\documentclass[pageno]{jpaper}

\usepackage[normalem]{ulem}

\usepackage{amsmath,amssymb,amsfonts}
\usepackage{textcomp}

\usepackage{wrapfig}

\usepackage{lipsum}
\usepackage[dvipsnames]{xcolor}
\usepackage{colortbl}
\usepackage{diagbox}

\usepackage{nicefrac}
\usepackage{booktabs} % for professional tables
\usepackage{multirow} % for professional tables
\usepackage{makecell} % for \multirowcell{}{}

\usepackage{pifont}% http://ctan.org/pkg/pifont

\usepackage[ruled,vlined,linesnumbered]{algorithm2e}

\usepackage{dblfloatfix}
\usepackage[bottom]{footmisc}

\usepackage{graphicx}
\usepackage{adjustbox}
\usepackage{bm}

\usepackage[para,online,flushleft]{threeparttable}

\usepackage[keeplastbox]{flushend}
\usepackage{xspace}

\usepackage{todonotes}

\newcommand{\eg}{\textit{e.g.}\xspace}
\newcommand{\ie}{\textit{i.e.}\xspace}

\newcommand*{\x}{\mathsf{x}\mskip1mu}

\newcommand{\squishitemize}{
 \begin{list}{$\bullet$}
  { \setlength{\itemsep}{-0pt}
     \setlength{\parsep}{0pt}
     \setlength{\topsep}{0pt}
     \setlength{\partopsep}{0pt}
     \setlength{\leftmargin}{-0.0em}
     \setlength{\labelwidth}{0.0em}
     \setlength{\labelsep}{0.1em} } }

\newcounter{Lcount}
\newcommand{\squishlist}{
    \begin{list}{\arabic{Lcount}. }
   { \usecounter{Lcount}
        \setlength{\itemsep}{-0pt}
        \setlength{\parsep}{3pt}
        \setlength{\topsep}{0pt}
        \setlength{\partopsep}{0pt}
        \setlength{\leftmargin}{2em}
        \setlength{\labelwidth}{1.5em}
        \setlength{\labelsep}{0.5em} } }

\newcommand{\squishend}{\end{list}}

\SetCommentSty{mycommfont}

\makeatletter
    \def\footnoterule{\kern-0\p@        % you can put other values to increase vertical space between rule and notes (just try out); difference between the values after "kern" is the width of the rule!
  \hrule \@width 1in \kern 0.1\p@}  % the in value is the length of the footnoterule
\makeatother

\usepackage{cite}
\definecolor{Mycolor2}{HTML}{007FFF}
\definecolor{Mycolor2}{HTML}{007FFF}
\newcommand\hl[1]{%
  \bgroup
  \hskip0pt\color{Mycolor2}%
  #1%
  \egroup
}

\definecolor{linkcolors}{HTML}{000000}
\usepackage[capitalise,noabbrev]{cleveref}
\crefformat{section}{\S#2#1#3} % see manual of cleveref, section 8.2.1
\crefformat{subsection}{\S#2#1#3}
\crefformat{subsubsection}{\S#2#1#3}

\title{\vspace{-20pt}LCP: A Low-Communication Parallelization Method\\for Fast Neural Network Inference in Image Recognition}
\author{\large Ramyad Hadidi, Bahar Asgari, Jiashen Cao, Younmin Bae, Da Eun Shim, Hyojong Kim, \\
  \large Sung-Kyu Lim, Michael S. Ryoo$^\dagger$, Hyesoon Kim\\
  \normalsize Georgia Institute of Technology, $^\dagger$Stony Brook University\vspace{-15pt}
  }
\begin{document}
\date{}
\maketitle
\thispagestyle{empty}

%%%%%% -- PAPER CONTENT STARTS-- %%%%%%%%

%---------------------------------
% \begin{abstract}
%
%
%
\textbf{\emph{Abstract ---}} \emph{Deep neural networks (DNNs) have inspired new studies in myriad edge applications with robots, autonomous agents, and Internet-of-things (IoT) devices. However, performing inference of DNNs in the edge is still a severe challenge, mainly because of the contradiction between the intensive resource requirements of DNNs and the tight resource availability in several edge domains. Further, as communication is costly, taking advantage of other available edge devices by using data- or model-parallelism methods is not an effective solution. To benefit from available compute resources with low communication overhead, we propose the first DNN parallelization method for reducing the communication overhead in a distributed system. We propose a low-communication parallelization (LCP) method in which models consist of several almost-independent and narrow branches. LCP offers close-to-minimum communication overhead with better distribution and parallelization opportunities while significantly reducing memory footprint and computation compared to data- and model-parallelism methods. We deploy LCP models on three distributed systems: AWS instances, Raspberry Pis, and PYNQ boards. We also evaluate the performance of LCP models on a customized hardware (tailored for low latency) implemented on a small edge FPGA and as a 16mW 0.107mm$^2$ ASIC @7nm chip. LCP models achieve a maximum and average speedups of $56\x$ and $7\x$, compared to the originals, which could be improved by up to an average speedup of $33\x$ by incorporating common optimizations such as pruning and quantization.}
% \end{abstract}

%%---------------------------------
\section{Introduction \& Motivation}
\label{sec:intro}
\noindent
%%% -----------------------------------------------------------------
%%% -----------------------------------------------------------------
%
%
The advancements of deep neural networks (DNNs) have made revolutionary changes in domains such as robotics~\cite{giu:guz16, pfe:sch17, corcoran2016mobile, veloso2012cobots, hadidi2018distributed}, unmanned aerial vehicles (UAVs)~\cite{sin:gan16, lu:li18, fu2019secure, mohamed2014service}, and Internet-of-things (IoT)~\cite{yao2018deep, sez:dog18, li2018learning, tran2017collaborative, hadidi2020towards, grieco2014iot}. In these domains, such as smart homes/cities/offices (\eg, connected cameras, gaming consoles, TVs, routers) or collaborative robots/drones (\eg, disaster relief~\cite{erdelj2017wireless, quaritsch2008collaborative, michael2014collaborative}, agriculture~\cite{bechar2016agricultural, anil2015revolutionizing}, mining~\cite{baudoin2010using}, construction~\cite{de2018distributed}, mapping~\cite{michael2014collaborative, golodetz2018collaborative}), (i) ensuring an acceptable accuracy is enough (\eg, detecting human sound in a disaster area with either 87\% or 90\% accuracy necessitates more investigation); (ii) the network of devices is standalone (\ie, Internet connection is not available/necessary); and (iii) the network has a unified ownership and hence communication among devices is not hazardous (\eg, IoTs at home, robots at warehouse). In such domains, executing inference in-the-edge could enable several features; however, performing the heavy inference computations locally is still a challenge. \emph{This paper enables performing DNN inference locally with efficient distribution and parallelization in the edge environments.}

%%% -----------------------------------------------------------------
%%% -----------------------------------------------------------------
%
The widespread approach to deal with the heavy inference computations of DNNs is to offload the requests and private data to high-performance servers of cloud providers~\cite{gupta2015discovering, li2015internet}. However, cloud-based offloading is not always available (\eg, no Internet access) and often relies on unreliable network latency. Furthermore, with the exponential increase in the number of edge devices~\cite{gartner-iot} and the scale of raw collected data, centralized cloud-based approaches might not scale~\cite{gubbi2013internet, zhan:ben15}. Privacy concerns~\cite{IotSurvey,gartner-iot-datacenter,lee:lee15} and personalization are other main driving forces for in-the-edge inference. However, the challenge is that performing DNN inference locally in the edge demands high compute and memory resources that contradict the energy, computational, and economical profile of edge devices~\cite{han:mao15, had:iiswc19}.

%%% -----------------------------------------------------------------
%%% -----------------------------------------------------------------
\noindent
\textbf{The Current Approach:}
The current approach for enabling local DNN inference while adhering to edge devices computational and economical profile is to locally distribute inference computations by taking advantage of the existing surrounding devices such as idle IoT devices. The distribution is based on data- or model-parallelism methods~\cite{kri:sut12, dean:cor12}. In data parallelism, the entire model is duplicated on each device for performing \emph{separate inferences}. Hence, the system needs several live and concurrent inputs to be efficient without real-time jitter. Simply put, data parallelism only increases throughput. In model parallelism, the model is divided and distributed across several devices for \emph{the same inference}.

\noindent
\textbf{The Key Challenge:}
The communication overhead and the inherent inter-layer data dependency limits effective parallelism. Therefore, an ideal parallelization method for edge devices, must minimize the communication overhead, while yielding low memory and computation footprints per node. However, none of the current distribution methods jointly reduce memory usage, computations, and communication (see Table~\ref{table:compare-intro}). \cref{sec:motive} presents a detailed description.

%%% -----------------------------------------------------------------
%%% -----------------------------------------------------------------
\noindent
\textbf{Our Solution:}
To address the aforementioned challenge, we propose a low-communication parallelization (LCP) method that enables the following:
\emph{\textbf{(i)} Reduces Communication:} LCP models replace a single, wide, and deep model with several narrow ones that only communicate for input and pre-final activations. Thus, their communication load is low with distributions (see Table~\ref{table:compare-intro}).
\emph{\textbf{(ii)} Reduces Compute \& Memory Footprints Per Node:} LCP models have fewer connections than those of the original ones, so their number of parameters and computational demands are also lower than those of the peer model-parallelism versions, shown in Table~\ref{table:compare-intro}.
\emph{\textbf{(iii)} Allows Inter-Layer Parallelism:} Narrow branches in LCP models are independent of each other, which enables inter-layer parallelism. This is in contrast to model parallelism that only allows intra-layer parallelism due to the single-chain dependency between consecutive layers.
\emph{\textbf{(iv)}: Recovers Accuracy with no Additional Parameters:} After splitting the model into branches, to recover a possible accuracy loss, LCP may slightly fatten the branches. However, since it reduces unnecessary communications, the overall parameters after fattening are still fewer than the original one.
%

%>>>>>>>>>>>>>>>>>>>>>>>
%>>>>>>>>>>>>>>>>>>>>>>>
%>>>>>>>>>>>>>>>>>>>>>>>
% >>>>>>>>>>>>>>>>>>>>>>>>>>
% >>>>>>>>>>>>>>>>>>>>>>>>>>
% >>>>>>>>>>>>>>>>>>>>>>>>>>
%
%
%
\definecolor{bad}{rgb}{0.980, 0.878, 0.892}
\definecolor{ideal}{rgb}{0.85, 0.99, 0.852}
\definecolor{ideal2}{rgb}{0.592, 0.945, 0.533}
\definecolor{good}{rgb}{0.992, 0.984, 0.886}
\definecolor{grayMy}{rgb}{0.968, 0.968, 0.968}

\renewcommand{\arraystretch}{0.90}
\begin{table}[!t]
\centering

\begin{adjustbox}{max width=\columnwidth}

\begin{threeparttable}
\centering  

\vspace{-4pt}
\caption{Methods for distributing inference computations. \vspace{-5pt}}

\scriptsize

\begin{tabular}{| c | c | c | c | c |}

      \hline
      \cellcolor{grayMy}
      & \cellcolor{grayMy}\textbf{Data} 
      & \cellcolor{grayMy} \textbf{Model}
      & \cellcolor{grayMy} 
      & \cellcolor{grayMy}  
      \\
      
      \cellcolor{grayMy}
      & \cellcolor{grayMy} \textbf{Parallelism}
      & \cellcolor{grayMy} \textbf{Parallelism}
      & \multirow{-2}{*}{\cellcolor{grayMy}\textbf{Target}}
      & \multirow{-2}{*}{\cellcolor{grayMy} \textbf{LCP}}
      \\
      \hline
      
      %------------
      \multicolumn{1}{|c|}{\cellcolor{grayMy} \textbf{Memory}}
      & \cellcolor{bad}  
      & \cellcolor{ideal} 
      & \cellcolor{ideal}  
      & \cellcolor{ideal2} 
      \\
      
      \multicolumn{1}{|c|}{\cellcolor{grayMy} \textbf{Per Device}}
      & \multirow{-2}{*}{\cellcolor{bad}DNN}
      & \multirow{-2}{*}{\cellcolor{ideal} $\frac{1}{n}$DNN}
      & \multirow{-2}{*}{\cellcolor{ideal} $\frac{1}{n}$DNN}
      & \multirow{-2}{*}{\cellcolor{ideal2} $\leq \frac{1}{n}$DNN}
      \\
      \hline
      
      %------------
       \multicolumn{1}{|c|}{\cellcolor{grayMy} \textbf{Communication}}
      & \cellcolor{ideal} 
      & \cellcolor{bad}  Intermediates
      & \cellcolor{ideal} 
      & \cellcolor{good} 
      \\
      
       \multicolumn{1}{|c|}{\cellcolor{grayMy} \textbf{Per Inference}}
      & \multirow{-2}{*}{\cellcolor{ideal}$\text{IN}/\text{OUT}$}
      & \cellcolor{bad} $+ \text{IN}/\text{OUT}$
      & \multirow{-2}{*}{\cellcolor{ideal} $\text{IN}/\text{OUT}$}
      & \multirow{-2}{*}{\cellcolor{good} $\approx\text{IN}/\text{OUT}$}
      \\
      
      \hline
      
      %------------
       \multicolumn{1}{|c|}{\cellcolor{grayMy} \textbf{Computation}}
      & \cellcolor{bad}
      & \cellcolor{ideal}
      & \cellcolor{ideal}
      & \cellcolor{ideal2}
      \\
      
       \multicolumn{1}{|c|}{\cellcolor{grayMy} \textbf{Per Device}}
      & \multirow{-2}{*}{\cellcolor{bad} DNN}
      & \multirow{-2}{*}{\cellcolor{ideal} $\frac{1}{n}$DNN}
      & \multirow{-2}{*}{\cellcolor{ideal} $\frac{1}{n}$DNN}
      & \multirow{-2}{*}{\cellcolor{ideal2} $\leq \frac{1}{n}$DNN}
      \\
      
      \hline
  
     %-------------------------------------------

\end{tabular}

\begin{tablenotes}[flushright]
    \item[] \hspace{50pt}DNN: Metrics associated with the entire model; $n$: Number of devices.
\end{tablenotes}

\label{table:compare-intro}

\end{threeparttable}

\end{adjustbox}

\vspace{-0pt}

\end{table}
\renewcommand{\arraystretch}{1}

%
%
%
% >>>>>>>>>>>>>>>>>>>>>>>>>>
% >>>>>>>>>>>>>>>>>>>>>>>>>>
% >>>>>>>>>>>>>>>>>>>>>>>>>>
%>>>>>>>>>>>>>>>>>>>>>>>
%>>>>>>>>>>>>>>>>>>>>>>>
%>>>>>>>>>>>>>>>>>>>>>>>

%%% -----------------------------------------------------------------
%%% -----------------------------------------------------------------

LCP is orthogonal and an addition to current techniques such as weight pruning~\cite{han:mao15} and quantization~\cite{gon:li14} that reduce the computational demand of models. LCP models offer distribution/parallelism opportunities for distributed computing, whereas current techniques apply accuracy/performance tradeoffs to single-node models. Such techniques can be applied to each branch of our method, as shown in \cref{sec:res:fpga}. Thus, LCP complement such techniques rather than compete with them.

%%% -----------------------------------------------------------------
%%% -----------------------------------------------------------------
\noindent
\textbf{Experiments Overview:}
\textbf{(1)} We generate and evaluate LCP models based on image-recognition DNNs on MNIST~\cite{mnist}, CIFAR10/100~\cite{krizhevsky2009learning}, Flower102~\cite{Flower102}, and ImageNet~\cite{rus:den15} datasets (total of 53 training results), covering all MLPerf~\cite{mattson2020mlperf} image-recognition models. \textbf{(2)} To evaluate the execution performance of our method, we conduct real-world implementations on three distributed systems with up to 10 Raspberry Pis (RPis), two PYNQ boards, and up to eight AWS instances. RPis are chosen because they represent the de facto choice for several robotic and edge use cases and they are readily available~\cite{vujovic2015raspberry, grimmett2015raspberry, millard2017pi, brand2018pidrone, wilson2016pheeno}. \textbf{(3)} We also evaluate the performance of LCP on customized hardware. Because, besides tailoring models based on hardware limitations, the architecture of hardware could be tailored to better achieve the goal of fast inference. To this end, we slightly modify the architecture of TPU~\cite{tpu} to make it latency-optimized  for edge applications rather than throughput-optimized, and implement it on a small Xilinx FPGA.  \textbf{(4)} To further investigate area and power efficiency of our tailored hardware for integrating with edge devices, we implement an ASIC chip in ASAP 7\,nm~\cite{clark2016asap7}.

\noindent
\textbf{Contributions:} Our contributions are as follows:
 \begin{list}{$\bullet$}
     {\setlength{\itemsep}{-1pt}
     \setlength{\parsep}{2pt}
     \setlength{\topsep}{0pt}
     \setlength{\partopsep}{0pt}
     \setlength{\leftmargin}{1.0em}
     \setlength{\labelwidth}{1.5em}
     \setlength{\labelsep}{0.5em}}
 \item We propose the first DNN parallelization method to reduce the communication overhead for distributed inference.
 \item We generate LCP models, with inter-layer parallelism for fast inference at small memory and computation footprints.
 \item We investigate the impact of hardware/software co-design on inference performance, by tailoring the hardware of TPU~\cite{tpu} for optimizing single-batch inference latency, and implement it on a small FPGA and as a tiny 0.107mm$^2$ low-power chip consuming only 16mW.
\end{list}

%%---------------------------------
\section{Challenges}
\label{sec:motive}
\noindent
\noindent
We first explain inevitable resource limitation for executing DNNs causing the single device Pareto frontier. Then, we summarize current distribution methods and their limitations, causing straggler problem and limited scope of parallelism.

\begin{figure}[h]
  \vspace{-10pt}
  \centering
  \includegraphics[width=1.0\linewidth]{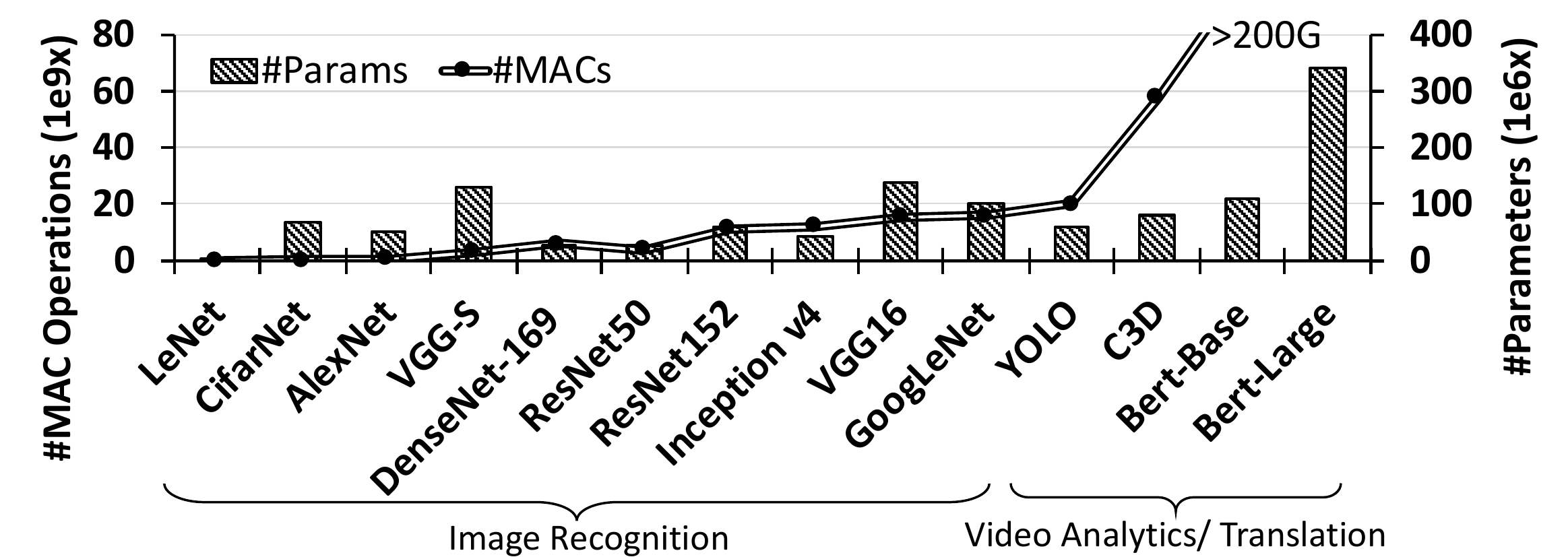}
  \vspace{-15pt}
  \caption{DNNs \#MAC operations/inference and parameters.}
  \vspace{-10pt}
  \label{fig:models-macs}
\end{figure}

%-------------------------------------
%-------------------------------------
%-------------------------------------
%
\noindent
\textbf{Resource Limitation \& Pareto Frontier:}
\label{sec:motive}
DNN models consist of several layers, the computations of which are based on custom weights that are learned during the training phase with back-propagation. In the inference, feed-forward computations are performed on batched inputs, and learned parameters stay constant. The most compute- and data-intensive layers~\cite{ven:ran17} are fully connected and convolution layers.\footnotemark~ Figure~\ref{fig:models-macs} shows the number of multiply-accumulate operations and parameter size in several DNN models. As shown, generally newer models encapsulate more parameters and perform more computations for better and more generalized feature understanding than their predecessors.
\emph{In short, this trend of modern models will inevitably surpass the capabilities of any resource-constrained device.}

\footnotetext{Since this paper focuses on visual models, we only introduced the layers in such models. For future work, we aim to include other types of DNNs. }

\begin{figure}[b]
  \vspace{-0pt}
  \centering
  \includegraphics[width=1.0\linewidth]{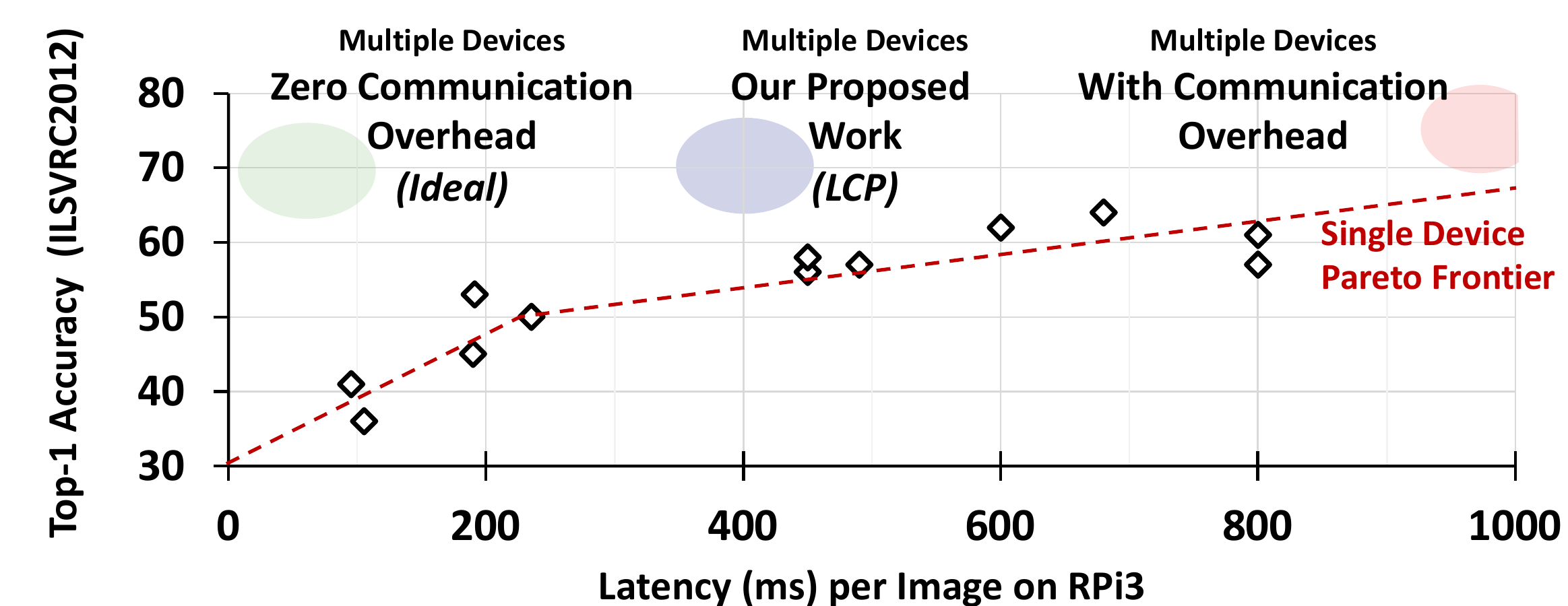}
  \vspace{-15pt}
  \caption{Latency-Accuracy Pareto Frontier -- Single device: Latency per image on RPi3 for ILSVRC models with the optimized platform-specific compilation ELL~\cite{ell} tool~\cite{defer-sysml-talk}. Multiple devices: Breaking the single device Pareto frontier, but with significant communication overhead.} 
  \vspace{-0pt}
  \label{fig:rpi-times}
\end{figure}
The capabilities of resource-constrained platforms are limited. Figure~\ref{fig:rpi-times} depicts latency per image using state-of-the-art image recognition models on RPi~\cite{defer-sysml-talk}. All implementations heavily utilize modern machine learning optimizations such as pruning~\cite{han:mao15}, quantization, low-precision inference~\cite{gon:li14, van:sen11,lin2016fixed}, and handcrafted models~\cite{ian:mos16}. Additionally, the models are highly optimized for ARMv8 architectures using the ELL compilation tool~\cite{ell}. However, achieving higher execution performance is impossible on a single device due to the Pareto frontier. As seen, the latency for high-accuracy models is longer than 400ms, and generally, latencies are longer than 100ms. In addition, the data shown in the figure is only for image-recognition models; DNNs in other domains are already surpassing these models in size and complexity. Fitting such an exponentially increasing computation on a single device, especially for edge devices, is a limiting factor for executing DNNs in the edge. 
\emph{In other words, even after applying all optimization techniques for DNNs, the single device Pareto frontier limits the widespread applicability of DNNs in several edge domains necessitating distribution and parallelization.}

%-------------------------------------
%-------------------------------------
%-------------------------------------
%
\begin{figure}[t]
  \vspace{-5pt}
  \centering
  \includegraphics[width=0.85\linewidth]{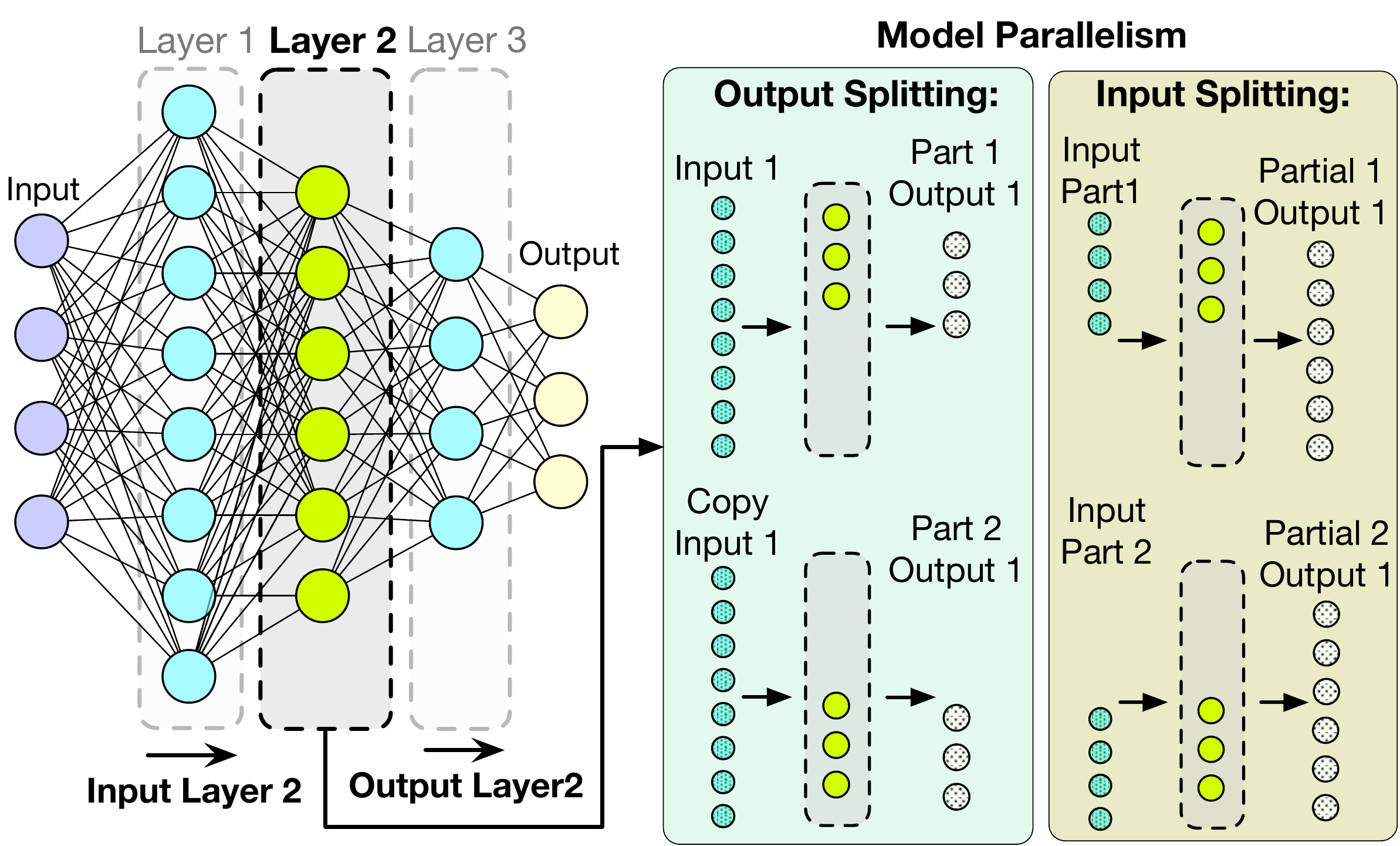}
  \vspace{-5pt}
  \caption{Model parallelism for a fully connected layer.}
  \label{fig:fc-why-not}
  \vspace{-5pt}
\end{figure}
\begin{figure}[b]
  \vspace{-5pt}
  \centering
  \includegraphics[width=0.90\linewidth]{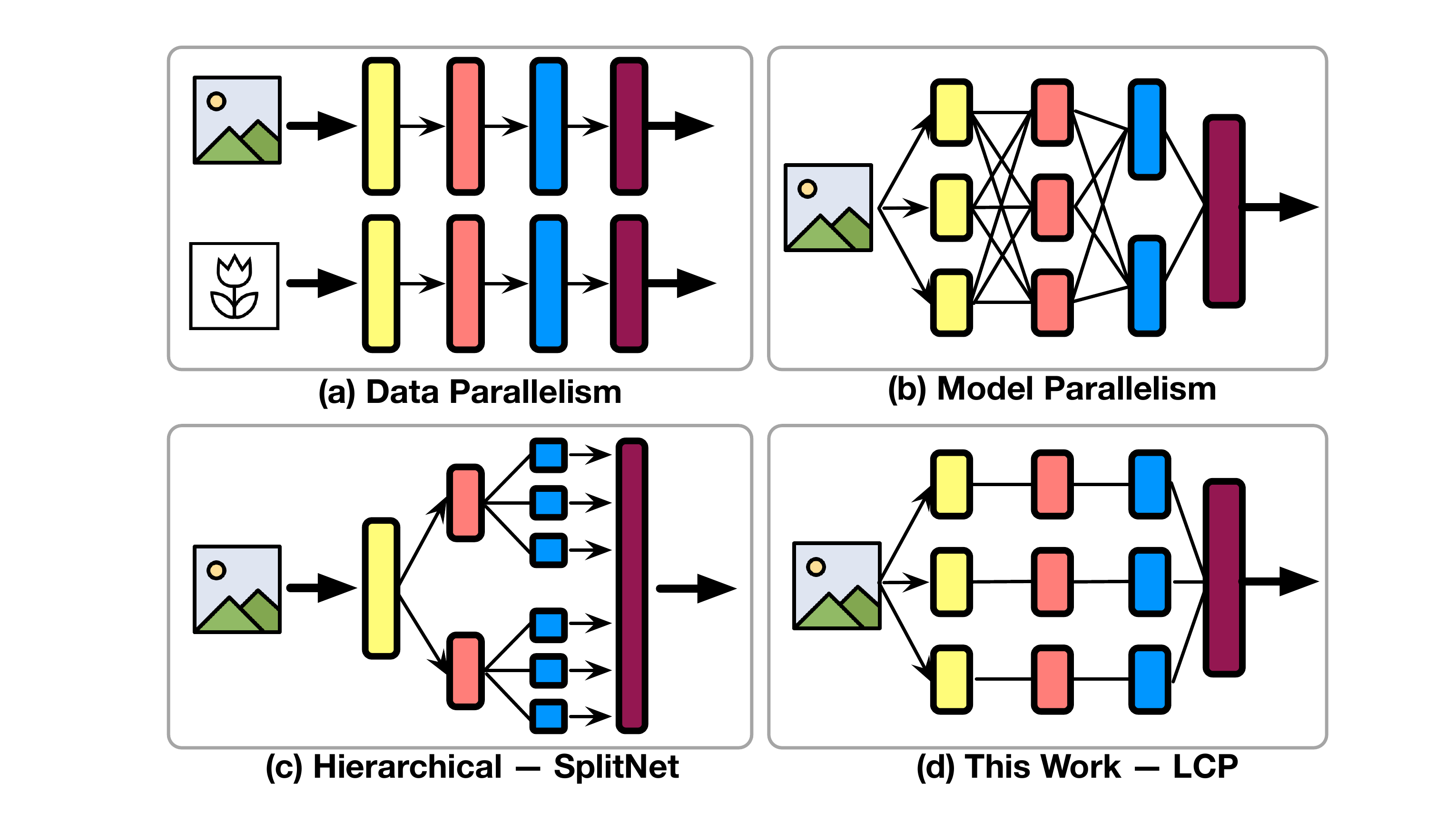}
  \vspace{-5pt}
  \caption{Overview of distribution/parallelism methods.}
  \label{fig:model-data}
  \vspace{-5pt}
\end{figure}

\noindent
\textbf{Current Distribution Methods:}
\label{sec:motive}
\textbf{(1)} Data parallelism (Figure~\ref{fig:model-data}a) parallelizes the computations of independent inputs~\cite{kri:sut12, dean:cor12}. Data parallelism does not apply to the edge because: It (i) serves several independent requests, the number of which is limited in an edge environment; (ii) does not reduce \emph{end-to-end latency} per inference and only increases throughput. Latency is important in several applications in the edge; and (iii) does not change the computation and memory footprints per node (Table~\ref{table:compare-intro}).

\textbf{(2)}
Model-parallelism (Figure~\ref{fig:model-data}b) divides the inference computations for the same request~\cite{kri:sut12, dean:cor12}. This method divides the computations within layer(s) while keeping dependencies intact. Depending on the type of layer, the dividing could take several forms. Figure~\ref{fig:fc-why-not} presents a simple example for distributing a fully connected (\texttt{fc}) layer, illustrating two extremes of model parallelism: Input and output splitting~\cite{hadidi2020towards}.  In output splitting, producing each output(s) is divided among the devices. In input splitting, the input is split and each device computes all parts of the output that are dependent on their received input. As shown in Figure~\ref{fig:fc-why-not}, each method has communication overhead (transmission of the input to all nodes or partial sums to a final node for summation). New model-parallelism methods is also crafted by mixing these two extremes, but they similarly suffer from the same discussed overhead. Several model-parallelism methods also exist for convolution layers by using matrix-matrix multiplication~\cite{chet:wool14, had:abu15}. Model parallelism does not change the interconnection of a model.
\emph{Hence, although model parallelism reduces the compute and memory footprint per node; the single-chain dependency between consecutive layers limits the parallelism scope within a single inference and causes communication overhead.}

\textbf{(3)}
SplitNet~\cite{kim2017splitnet}, shown in Figure~\ref{fig:model-data}c, gradually splits the model in a tree-structured style \emph{manually} based on the dataset semantics, extracted in intermediate to final layers. Therefore, SplitNet (i) splits only intermediate to final layers, (ii) is invariant to the number devices, (iii) creates imbalanced workload because of its dependency on semantics, (iv) results in tree-style connections, incurring high communication overhead, and (v) enforces a new splitting when dataset changes.

%-------------------------------------
%-------------------------------------
%-------------------------------------

%
\begin{figure}[t]
  \vspace{-5pt}
  \centering
  \includegraphics[width=0.95\linewidth]{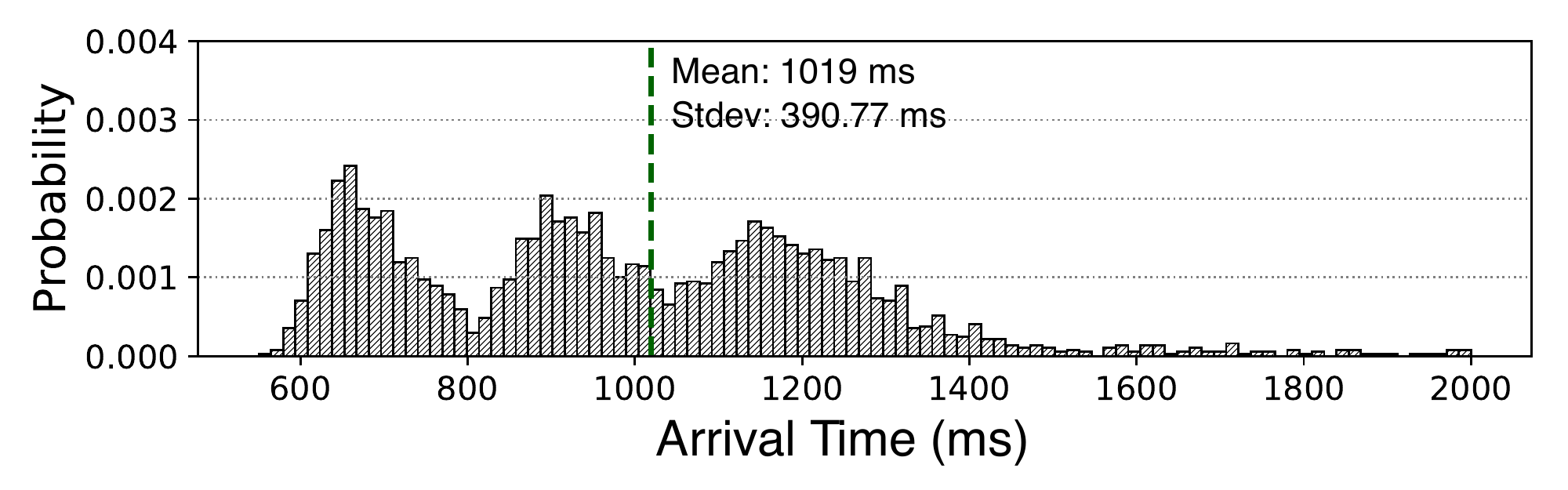}
  \vspace{-8pt}
  \caption{Histogram of prediction latencies on a six RPi system executing AlexNet with model parallelism (\cref{sec:res:rpi}).}
  \vspace{-5pt}
  \label{fig:alexnet-hist}
\end{figure}
\begin{figure}[b]
  \vspace{-0pt}
  \centering
  \includegraphics[width=0.90\linewidth]{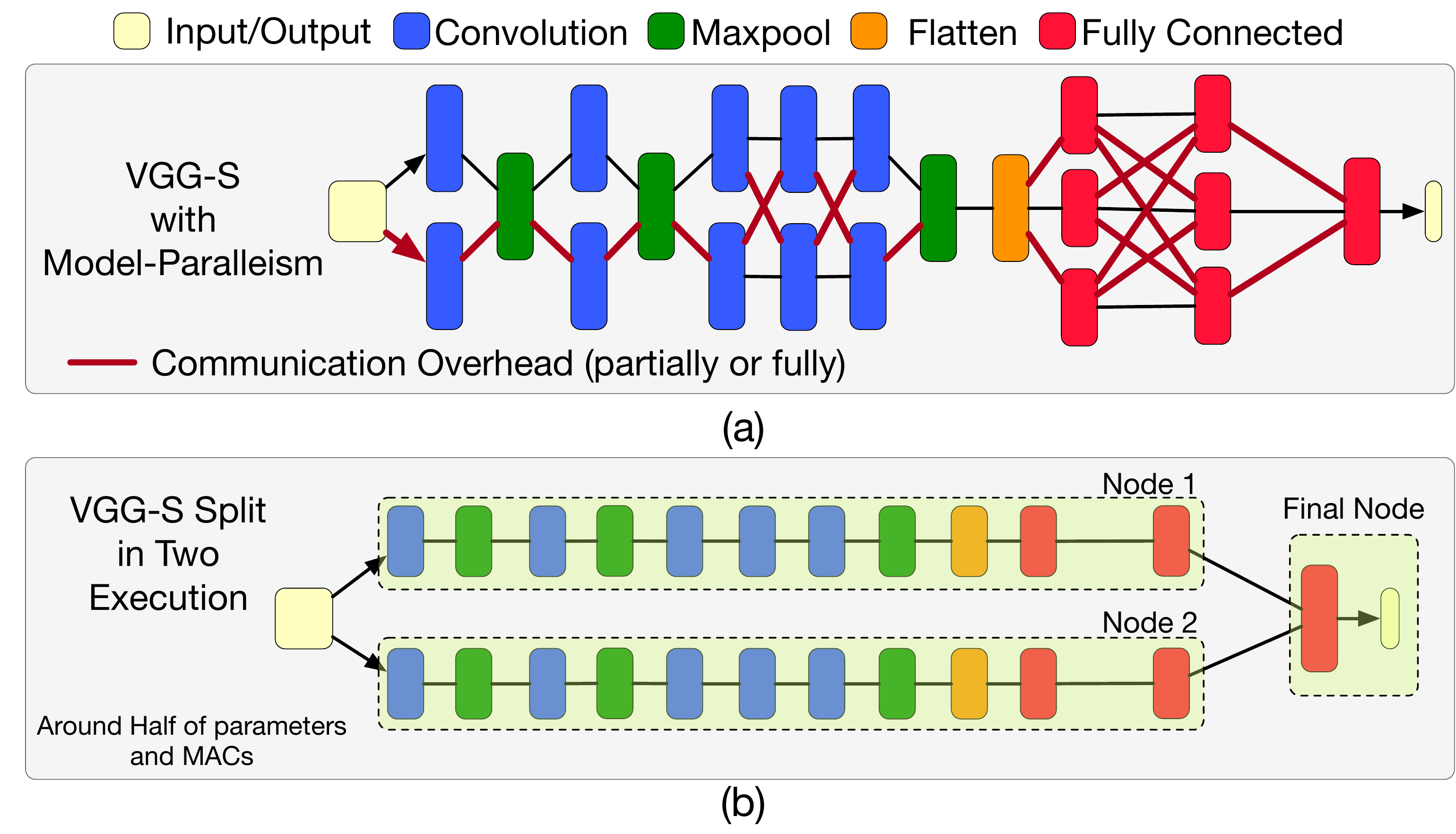}
  \vspace{-5pt}
  \caption{VGG-S (a) model parallelism and (b) LCP versions.}
  \label{fig:vgg_s}
  \vspace{-5pt}
\end{figure}

\noindent
\textbf{Communication Overhead \& Limited Parallelism:}
\label{sec:motive}
Current distribution methods have a high communication overhead and limited scope of parallelism which stems from the single-chain dependency between consecutive layers. High communication induces the straggler problem, in which a system is lagged by its slowest node. Specifically, since edge devices usually use a wireless network, the latency deviations are high. As an example, Figure~\ref{fig:alexnet-hist} depicts the histogram of prediction latencies on a distributed IoT system consisting of six RPis executing AlexNet with model parallelism. The computing time is bounded to 500ms, but the average delay is $\approx\!2\x$ longer (and $\approx\!4\x$ for tail latency). To gain perspective, Figure~\ref{fig:vgg_s}a shows VGG-S with model parallelism and its communication overhead. As seen, dependencies enforce a strongly interconnected network among the nodes. Although several techniques such as compression could alleviate the cost of communication, still the number of connections remains constant.
\emph{Therefore, an ideal distribution method for edge devices besides yielding low memory and computation footprints per node must consider communication overhead.}

The single-chain dependency between consecutive layers limits the available parallelism that could be harvested by the aforementioned methods. The limitation is that after the computations of a single/few layer(s) are done, the intermediate results must be merged before being forwarded to the next layer. Such merging acts as a global barrier, which similar to parallel programming, limits the gained performance speedup. \emph{In summary, with parallel execution on multiple devices, ideally, we could pass the frontier in Figure~\ref{fig:rpi-times}. However current distribution methods are limited by the communication overhead and the inherent inter-layer data dependency. The next section proposes LCP models, which significantly reduce communication and allow inter-layer parallelism.}

%%---------------------------------
\section{LCP For Fast Inference}
\label{sec:ETP}
\noindent
To address challenges, we propose LCP method, which replaces a single, wide, and deep model with several narrow branches that only communicate for input and pre-final activation (Figure~\ref{fig:model-data}d).
Figure~\ref{fig:vgg_s}b shows an example of a two-branch LCP model for VGG-S.
This section first explains the design procedure  of LCP models and discusses their key features enabling low-communication parallelization. The second part focuses on tailoring a systolic architecture for edge computing.

%-----------------------------------
%-----------------------------------
%-----------------------------------
\subsection{Tailoring Models}
\label{sec:ETP:etp}

\begin{figure}[b]
  \vspace{-0pt}
  \centering
  \includegraphics[width=0.90\linewidth]{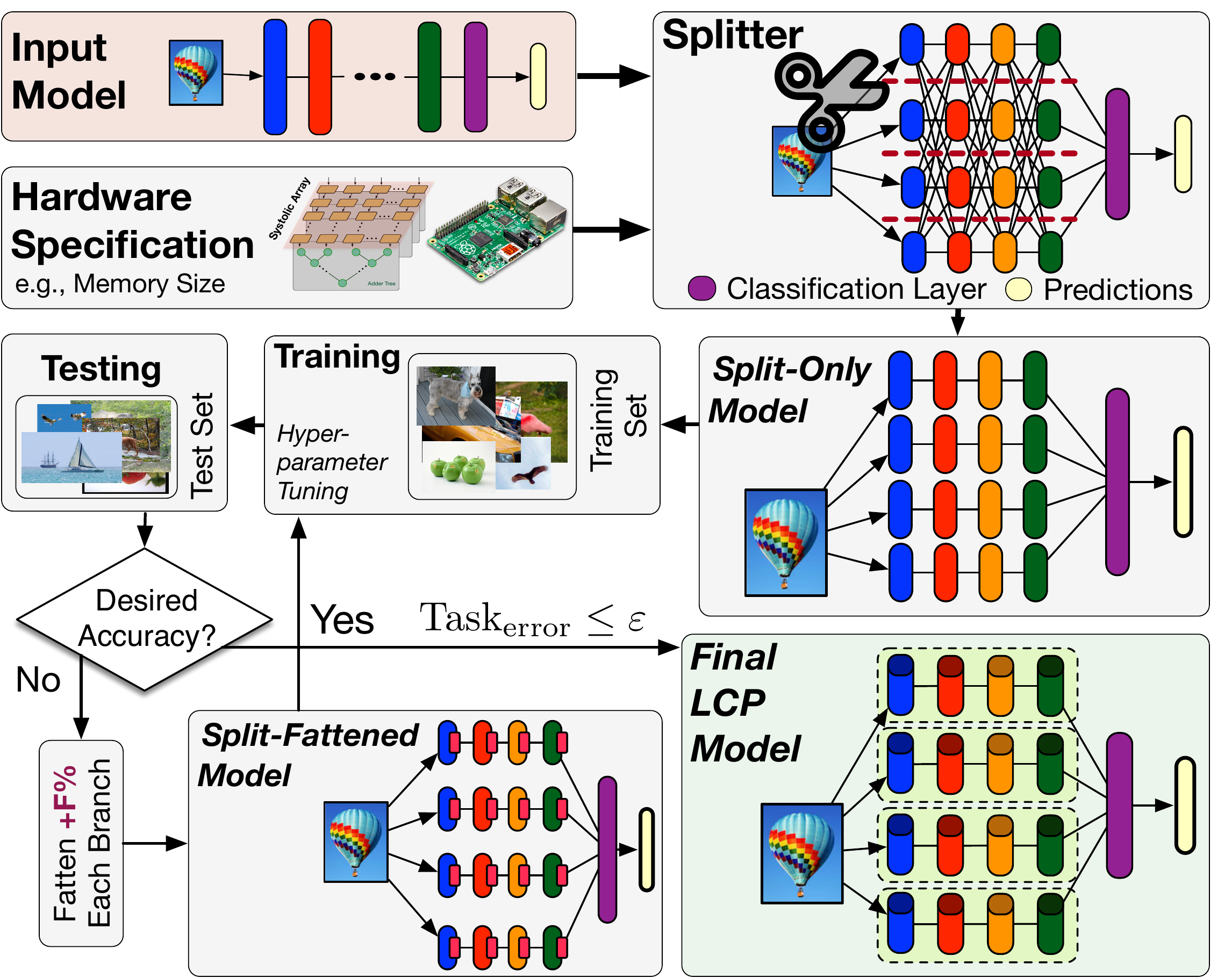}
  \vspace{-5pt}
  \caption{Design Procedure of LCP models.}
  \label{fig:proc}
  \vspace{-5pt}
\end{figure}

\noindent
\textbf{Design Procedure:}
\vspace{-0pt}
Figure~\ref{fig:proc} describes the design procedure of LCP models. 
We start by inputting the DNN model and its per-layer memory and computation footprints. Similarly, we input the specification of the hardware, such as memory size, computation capability, and any overhead associated with executing a DNN on our hardware. For instance, several DNN frameworks have a memory overhead because of the framework. A splitter procedure, described in Procedure~\ref{algo:split}, in a while loop, splits the model, cuts the connection, and measures the approximate footprints of each branch. The $\text{Division}_\text{Factor}$, a hyperparameter, defines the granularity of division/splitting. Here, we assume the $\text{Division}_\text{Factor}$ of two, but any number is viable. The loop exits when a single branch is fitted on a device (both memory and computation wise). If the number of devices is fewer than the number of branches, the execution is still possible, but will be inefficient. Then, we remove non-branch connections in a simple operation that keeps only one connection per layer. 
The derived model from the splitter is the \emph{split-only model}. By training the split-only model and testing it, we measure its accuracy. The split-only models have fewer parameters and MAC operations than the original models (see Table~\ref{table:models-vision}) in total. Hence, after distribution, each branch has less computation and memory footprint than its model-parallelism version.

As a result of fewer number of parameters and removing several connections, a slight accuracy drop in split-only LCP models is expected. Depending on the accuracy requirement of the task, we either fatten each branch by $F\%$, a hyperparameter, or output the resulted model. We assumed a maximum of 3\% bound for $\text{Task}_\text{error}$. Fattening each branch by $F\%$ is done by increasing the number of channels and output features of convolution and fully connected layers of the split-only model, respectively. Note that theses new \emph{split-fattened models} are fattened within each branch. Thus, even with a high fattening percentage, still they have fewer parameters and MAC operations than the original model  (see Table~\ref{table:models-imagenet-fat}). When the accuracy is in the acceptable error range for our task, $\text{Task}_\text{error}$, we output the model architecture and its weights. It is expected that with similar number of parameters after fattening, LCP models achieve the same level of accuracy~\cite{xie2019exploring}. We showcase LCP models in \cref{sec:res:etp} covering MLPerf~\cite{mattson2020mlperf}.

\makeatletter
\newcommand{\removelatexerror}{\let\@latex@error\@gobble}
\makeatother

\begin{table}[t]
\vspace{-5pt}
\noindent
\begin{center}
\begin{minipage}{0.95\linewidth}

\removelatexerror
\begin{algorithm}[H]

\SetAlgorithmName{Procedure}

  \scriptsize
  \SetKwInOut{Input}{Input}
  \SetKwInOut{Output}{Output}
  \SetKwFunction{split}{Split}
  \SetKwFunction{remove}{RemoveNonBranchConnections}
  \SetKwFunction{encodere}{CsrRleRepeat}
  \SetKwData{division}{Division$_\text{factor}$}
  \SetKwArray{dnn}{DNN}
  \SetKwArray{layer}{layer}
  \SetKwArray{layerw}{layer.width}
  \SetKwArray{dnnmem}{DNN$_\text{Mem}$}
  \SetKwArray{dnnmac}{DNN$_\text{MAC}$}
  \SetKwArray{devmem}{Dev$_\text{Mem}$}
  \SetKwArray{devmac}{Dev$_\text{MAC}$}
  \SetKwArray{mem}{Mem$_\text{fit}$}
  \SetKwArray{mac}{MAC$_\text{Mac}$}
  \SetKw{and}{and}
  \SetKwProg{myproc}{}{}{}
  
   \caption{LCP Splitter {\small(in Figure~\ref{fig:proc})}}
   \label{algo:split}

    \Input{\textit{\dnn}: Layer configurations $[0:n]$\\
           \textit{\dnnmem}, \textit{\dnnmac}: DNN memory and computational footprints\\
           \textit{\division}: Division Factor for splitting\\
           \textit{\devmem}, \textit{\devmac}: Hardware specification
    }
    \Output{ \textit{\dnn}: Layer configurations $[1:n]$ }
    \myproc{\split{\dnn, \dnnmem, \dnnmac, \division, \devmem, \devmac}}{
      \mem $\gets$ 0; \mac $\gets$ 0;\\
      \While{not \mem \and not \mac}{
        \mem $\gets$ \dnnmem $<$ \devmem \\ 
        \mac $\gets$ \dnnmac $<$ \devmac \\ 
        
        \For{\layer $[0..n-1]$ in \dnn} {
          \layerw $\gets$ \layerw / \division
        }
        \remove{\dnn}

      }
      \Return $<$\dnn$\!\!>$
      
      }
\end{algorithm}

\end{minipage}
\end{center}
\vspace{-17pt}
\end{table}

%-----------------------------------
%-----------------------------------
%-----------------------------------
\noindent
\textbf{Key Features of LCP Models:}
LCP models are designed by considering their underlying computation domain and have the following key features to address the challenges discussed in \cref{sec:motive}:
%\squishitemize
%\item
{\textbf{(1)}} LCP models only communicate for input and pre-final activation. Therefore, they significantly reduce communication overhead in a distributed system. Additionally, the low communication load per inference helps with the straggler problem. This is in contrast to model parallelism, which highly depends on communication among all the intermediate layers;
%\item
{\textbf{(2)}} LCP models split the size of a layer, so the total parameter size and computation complexity of the model are reduced. Therefore, they require fewer parameter sizes, less computation complexity, and no communication between the nodes for intermediate layers. These lower memory and computation footprints allow edge devices to efficiently operate within their limited resources (\eg, no swap space activities due to limited memory);
%\item
{\textbf{(3)}} LCP models replace the original wide model with several narrow and independent branches. Since the computations of branches are not dependent, in contrast to the single-chain of dependency in the original model, the scope of parallelism is not limited with each layer anymore. In other words, LCP models go beyond intra-layer parallelism.
%\squishend

\noindent
\begin{figure}[b]
  \vspace{-5pt}
  \centering
  \includegraphics[width=1.0\linewidth]{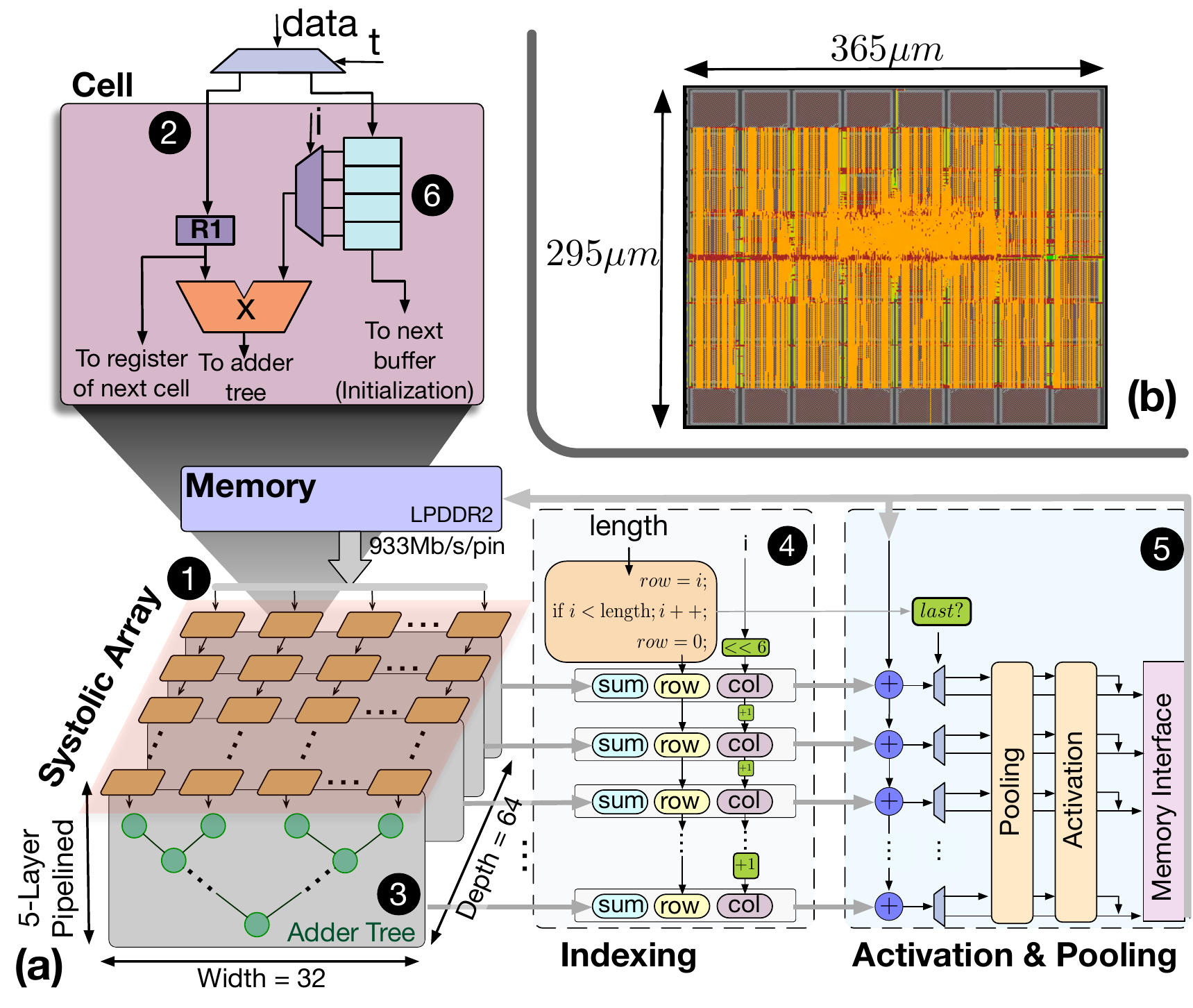}
  \vspace{-15pt}
  \caption{Details of Tailored Hardware for Edge: (a) Microarchitecture overview, and (b) Layout of ASIC design at 7nm.}
  \label{fig:micro}
  \vspace{-5pt}
\end{figure}
%
%
%

%-----------------------------------
%-----------------------------------
%-----------------------------------
\vspace{-10pt}
\subsection{Tailoring Hardware}
\label{sec:ETP:tailoring_hw}
% \noindent
% \textbf{Current Deep Learning Accelerators:}
Last section showed how we enable fast inference under resource constraints and at costly communication, by proposing a low-communication parallelization method that results in several narrow models. To further achieve the goal of fast inference and recognize the potential, the hardware can also be tailored. Recently, several popular tailored hardware designs for DNNs~\cite{Ese, eyeriss, shidiannao, tpu, dean2017machine, asgari2019eridanus} including TPU~\cite{tpu} use systolic arrays~\cite{kung1982systolic} that offer a high degree of concurrent processing through a dataflow compute arrays hence providing high \textit{throughput}. In the edge applications, however, the main goal is \textit{reducing single-batch inference latency}, rather than high throughput solely. This section introduces our microarchitecture (Figure~\ref{fig:micro}a), an example of tailoring and simplifying the architecture of TPU to be implemented on small FPGAs or be fabricated as tiny (i.e., 0.107~mm$^2$ as shown in Figure~\ref{fig:micro}b) low-power chips to be integrated with edge devices.

%
%
%-----------------------------------
%-----------------------------------
%-----------------------------------
% \subsection{Microarchitecture Details}
\label{sec:ETP:micro}

% \noindent
% \textbf{Overview:}
Figure~\ref{fig:micro}a illustrates our tailored microarchitecture that similar to TPU, comprises a weight-stationary systolic array~\cite{kung1982systolic} for implementing matrix-matrix multiplication. The systolic array cells are organized in a 32$\x$64 array \ding{182}. To reduce the number of connections, only the first row of the systolic array is connected to the memory \ding{182}. Moreover, each cell of the first row is only connected to one data stream line \ding{183}. Based on the type of an operand ($t$), streaming data is used for either initialization or for processing. Since the width of the systolic array is 32, a heuristic algorithm partitions the streaming (i.e., non-stationary operand) into blocks of 32 width and arbitrary length, and splits the stationary operand, into 32 $\times$ 64 blocks. To assist the smooth streaming of data from memory to the systolic array, we map these blocks along with their indices ($i$), type (stationary/non-stationary), and length to sequential memory addresses.
We implement our 32$\times$64 systolic array connected to \emph{LPDDR2} memory with the data rate of 933Mb/s/pin @466\,MHz~\cite{lpddr2}, which gives a bandwidth of 3.7~GB/s. Other packaging options with higher memory bandwidths are also feasible. However, seeking a fair comparison with RPi3s, we choose this memory technology. The maximum data reuse rate of our design is 64\,OPs/Byte, which leads to a peak throughput of 217.6\,GOPs/s.
The following explains three main modifications we made to this systolic architecture, to achieve our goal of \textit{reducing single-batch latency}.

%
% adder trees
% adder tree pipeline/depth
% discussion about depth and fine-grained parallelism
%
%-------------------------
\noindent
\textbf{(1) Adder Trees:}
Instead of MAC-based systolic arrays, we separate adders from multiplications by integrating adder trees, the well established components for DNN accelerators~\cite{zhang2015optimizing, zhang2018dnnbuilder, yu2019opu}, into systolic arrays architecture. Each cell of our systolic array is a \textit{multiplier} with two integer operands, one stationary and one streaming (\texttt{R1}). Each row of the multiplier array is connected to an adder tree~\ding{184}, pipelined in five ($log_2{32}$) stages. Adder trees reduce the result of multiplications into a single integer, which then contributes to creating an output element. The structure of the multiplier array connected to the adder trees reduces latency from O(n) to O(log(n)).

%
% indexing
% start and end signals
% activation/pooling
%
%-------------------------
\noindent
\textbf{(2) Simple Indexing Logic:}
We use a data-driven execution model, in which data is pushed by the memory to the systolic array, triggered by the arrival of data. During execution, for each element, the indexing logic (\ding{185}) generates the appropriate row and column indices of the element using the index ($i$) of a block and its length to accompany the result. The row and column indices will later be used by the memory interface to write the results to physical locations in memory. By comparing the length and index ($i$), the end of the operations in the current layer is detected. The end of the current layer signals the start of activation and pooling functions (\ding{186}) for that layer.

%
%
% Why buffers at Stationary
%
%-------------------------
\noindent
\textbf{(3) Buffering Stationary Operands:}
The stationary operands are often larger than the depth of the systolic array. In such cases, we have to partition a multiplication into several small operations that share a non-stationary operand, but have distinct stationary operands. To avoid multiple loads of stationary registers, we choose to integrate a buffer (\ding{187}) for stationary operands at each cell. As a result, the design serves requests with lower latency. Moreover, since each branch of the model has several layers, integrating these buffers allows fast context switching without the overhead of reloading the stationary operands. These buffers are connected in a column of cells, similar to streaming registers (\texttt{R1})s. During the initialization, stationary operands are poured into these connected buffers to fill them by utilizing the connections between them.

%%---------------------------------
\section{Experimental Studies}
\label{sec:res}
\noindent
This section shares our experimental results for generating LCP models and then their full-system implementation on RPi, TVM~\cite{chen2018tvm} on PYNQ boards, and AWS servers. Finally, we evaluated our hardware with edge FPGA implementation, and ASIC chip design. At the start of each subsection, the setup of related experiments is provided.

%
%-----------------------------
%-----------------------------
%-----------------------------
\subsection{Generating LCP Models}
\label{sec:res:etp}
%
%----------------------
\noindent
\textbf{Training Specifications:}
We train all the models, including the original model, from scratch to conduct a fair comparison. Normalization~\cite{iof:sze15} layers are included. The training is done with an exponential learning rate with a decay factor of $0.94$, initial learning rate $1\mathrm{e}{-2}$, number of epoch per decay of two or 10, a dropout rate of $50\%$, and L2 regularization with weight decay of $5\mathrm{e}{-4}$. We use ADAM optimizer~\cite{kingma2014adam} with $\beta_1=0.9$ and $\beta_2=0.99$. All biases are initialized to zeros and all weights are initialized with a normal distribution of mean 0 and a standard deviation of $4\mathrm{e}{-2}$. All of our models are trained until the loss is flattened or least for 12 epochs. Test and accuracy measurements are done on at least 10\% of datasets that have never been used in training to provide an unbiased evaluation of the model. For LCP, the $\text{Division}_\text{Factor}$, $F$, and $\varepsilon$, are 2\%, 10\%, and $\approx\!3\%$, respectively.

%----------------------
\noindent
\textbf{Datasets:}
We use the following datasets: (1) MNIST~\cite{mnist}, which contains 70k grayscale handwritten 28x28 images in 10 classes; (2) CIFAR10~\cite{krizhevsky2009learning}, which contains 60k colored 32x32 images in 10 classes; (3) CIFAR100~\cite{krizhevsky2009learning}, which contains 60k colored 32x32 images in 100 classes; (4) Flower102~\cite{Flower102}, which contains 16,378 colored 224x224 images of flowers in 102 classes; and (5) ImageNet~\cite{rus:den15}, which contains 1.33\,M colored 224$\x$224 images in 1000 classes.

\noindent
\textbf{Models:}
We use the representative model for each dataset, LeNet~\cite{lenet}, LeNet-FC~\cite{lenet}, VGG-S~\cite{sim:zis14-deep}, CifarNet~\cite{krizhevsky2009learning}, VGG16~\cite{sim:zis14-deep}, AlexNetv2~\cite{krizhevsky2014one}, ResNet-18/50~\cite{he:zha16}, and MobileNet~\cite{howard2017mobilenets}. We cover all image-recognition models in MLPerf. In total, for brevity, we only report 53 instances of training results to show LCP extensibility using five datasets and nine models. Our additional results (not reported) with ResNet-34, DenseNet~\cite{condensenet}, and DarkNet19~\cite{darknet13} confirms extendibility. Simple sequential DNNs serve as a basis to confirm our method, while ResNets and MobileNet showcase LCP with modern models.

\begin{figure}[b]
  \vspace{-0pt}
  \centering
  \includegraphics[width=1.0\linewidth]{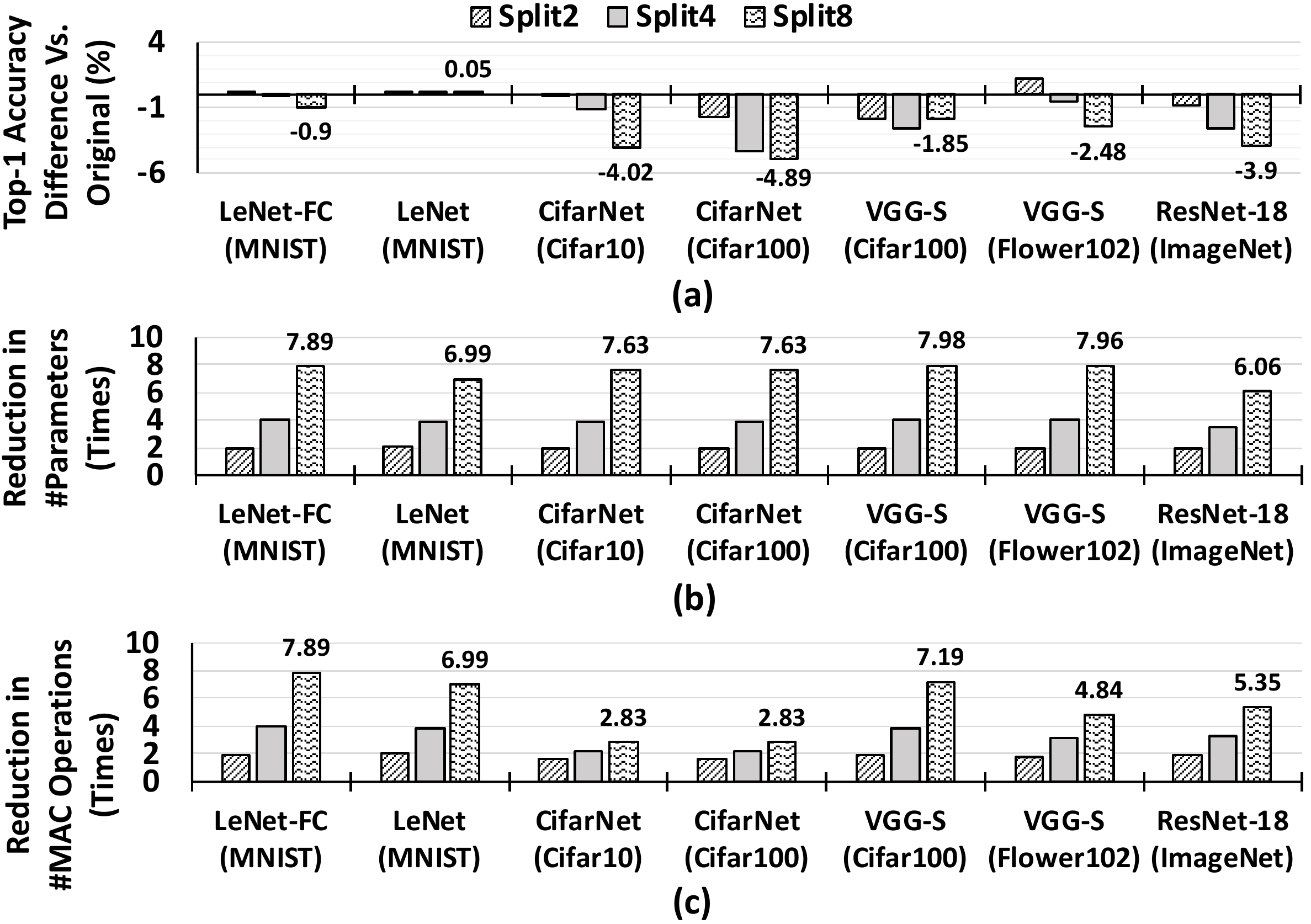}
  \vspace{-10pt}
  \caption{Split-Only Models: (a) Accuracy, (b) reduction in the number of parameters, and (c) reduction in the number of MAC operations in comparison with the original model.}
  \label{fig:acc-vision}
  \vspace{-0pt}
\end{figure}

%
%-----------------------------
\noindent
\textbf{Split-Only Models:}
For split-only models, we use $\text{Division}_\text{Factor}$ of two, which results in models with two, four, and eight branches. Except the width, defined as output features in fully connected layers and the number of output channels (i.e., filters) in convolution layers, the rest of the parameters are similar to the original model as Splitter Procedure~\ref{algo:split} only touches widths.
Table~\ref{table:models-vision} lists the training results. Figure~\ref{fig:acc-vision}a illustrates the accuracy difference of our models, shown in Table~\ref{table:models-vision}. As shown, the maximum accuracy drop is around 5\% for CifarNet. Note that this accuracy drop occurs when we reduced the parameter size of our model extensively (around $\nicefrac{1}{8}$). Figure~\ref{fig:acc-vision}b and c show reduction in the number of parameters and computation compared with the original DNN model; as seen, each split reduces both by about $\text{split}_{factor}$ times. This is because each convolution and fully connected layer in the split version create fewer outputs; therefore, the next layer requires fewer parameters. In the next section, we restore the accuracy of LCP models with split-fattened models.

\renewcommand{\arraystretch}{0.70}
\begin{table}[t]

\centering

\begin{adjustbox}{width=\columnwidth}

\begin{threeparttable}
\vspace{-0pt}
\centering

\caption{Results of split-only LCP models.\vspace{-5pt}}

\scriptsize

\vspace{-5pt}
\begin{tabular}{c || c | c | c | c | c }
  \toprule
  \multirow{2}{*}{\textbf{Model Name}} 
      & \multirow{2}{*}{\textbf{Dataset}} 
      & \multirow{2}{*}{\textbf{Layers$^\dagger$}} 
      & {\textbf{Top-1}}
      & \textbf{\#} 
      & \textbf{\# MAC} \\
   
      &      
      & 
      & {\textbf{Accuracy}}
      & \textbf{Param}
      & \textbf{Opr.}\\
  \midrule

     \textbf{LeNet-FC*}
     
     & \textbf{MNIST} 
     & \textbf{3fc} 
     & \textbf{97.95}
     & \textbf{266.6k}
     & \textbf{266.2k}
     
    %  \\    
    %  LeNet-FC-split2
     
    %  & MNIST
    %  & 10
    %  & 28$\x$28    
    %  & 2 
    %  & 5fc 
    %  & 98.20
    %  & 99.99
    %  & 134.0k
    %  & 134.1k
     
    %  \\     
    %  LeNet-FC-split4
     
    %  & MNIST
    %  & 10 
    %  & 28$\x$28   
    %  & 4 
    %  & 9fc
    %  & 97.84
    %  & 99.98
    %  & 67.7k
    %  & 67.1k
     
    %  \\
    %  LeNet-FC-split8
     
    %  & MNIST 
    %  & 10
    %  & 28$\x$28   
    %  & 8 
    %  & 17fc 
    %  & 97.05
    %  & 99.95
    %  & 33.9k
    %  & 33.8k
     
     \\
     \midrule
     %-------------------------------------------
     \textbf{LeNet}
     
     & \textbf{MNIST}
     & \textbf{2fc-3c-2p}
     & \textbf{98.76}  
     & \textbf{61.7k}
     & \textbf{61.5k}
     
    \\
      LeNet-split2
     
      & MNIST
      & 3fc-6c-4p 
      & 98.86 
      & 31.5k
      & 30.5k
     
      \\
      LeNet-split4
     
      & MNIST
      & 5fc-12c-8p 
      & 98.93 
      & 16.1k
      & 16.0k
     
      \\
      LeNet-split8
     
      & MNIST
      & 9fc-24c-16p 
      & 98.81
      & 8.8k
      & 8.5k
     
     \\
     \midrule

     %-------------------------------------------
     \textbf{CifarNet*}
     
     & \textbf{Cifar10} 
     & \textbf{2fc-2c-2p-2n-1d}
     & \textbf{80.72}
     & \textbf{797.97k}
     & \textbf{14.79M}
     
%     \\
%     CifarNet-split2
%     
%     & Cifar10
%     & 5fc-4c-4p-4n-2d 
%     & 80.63 
%     & 401.75k
%     & 9.32M
%     
%     \\
%     CifarNet-split4
%     
%     & Cifar10
%     & 9fc-8c-8p-8n-4d
%     & 79.53
%     & 203.64k
%     & 6.59M
%     
%     \\
%     CifarNet-split8
%     
%     & Cifar10
%     & 17fc-16c-16p-16n-8d
%     & 76.70
%     & 104.6k
%     & 5.22M

     \\
     \midrule
     
     %-------------------------------------------
     \textbf{CifarNet}
     
     & \textbf{Cifar100} 
     & \textbf{2fc-2c-2p-2n-1d}
     & \textbf{52.87}
     & \textbf{815.34k}
     & \textbf{14.81M}
     
     \\
     CifarNet-split2
     
     & Cifar100
     & 5fc-4c-4p-4n-2d 
     & 51.22
     & 410.48k
     & 9.33M
     
     \\
     CifarNet-split4
     
     & Cifar100
     & 9fc-8c-8p-8n-4d
     & 48.48
     & 208.05k
     & 6.59M
     
     \\
     CifarNet-split8
     
     & Cifar100
     & 17fc-16c-16p-16n-8d
     & 47.98
     & 106.85k
     & 5.23M

     \\
     \midrule
     %-------------------------------------------
     \textbf{VGG-S*}
     
     & \textbf{Cifar100}
     & \textbf{3fc-5c-2p-1n-2d }
     & \textbf{50.33}
     & \textbf{76.15M}
     & \textbf{154.09M}
     
%     \\
%     VGG-S-split2
%     
%     & Cifar100  
%     & 5fc-10c-4p-2n-4d  
%     & 48.53  
%     & 38.09M
%     & 78.28M
%     
%     \\
%     VGG-S-split4
%     
%     & Cifar100
%     & 9fc-20c-8p-4n-8d
%     & 47.72    
%     & 19.06M
%     & 40.37M
%     
%     \\
%     VGG-S-split8
%     
%     & Cifar100
%     & 17fc-40c-16p-8n-16d 
%     & 48.48
%     & 9.54M
%     & 21.42M

     \\
     \midrule
     %-------------------------------------------
     \textbf{VGG-S}
     
     & \textbf{Flower102}
     & \textbf{3fc-5c-3p-1n-2d} 
     & \textbf{88.14}
     & \textbf{60.79M}
     & \textbf{1.85G}
     
     \\
     VGG-S-split2
     
     & Flower102  
     & 5fc-10c-6p-2n-4d  
     & 89.31  
     & 30.50M
     & 1.01G
     
     \\
     VGG-S-split4
     
     & Flower102
     & 9fc-20c-12p-4n-8d
     & 87.55    
     & 15.26M
     & 591.65M
     
     \\
     VGG-S-split8
     
     & Flower102
     & 17fc-40c-24p-8n-16d 
     & 85.66
     & 7.64M 
     & 382.51M

     \\
     \midrule
     %-------------------------------------------
     \textbf{ResNet-18}
     
     & \textbf{ImageNet}
     & \textbf{18c-2p-17n} 
     & \textbf{70.68}
     & \textbf{11.69M}
     & \textbf{1.82G}
     
     \\
     ResNet-18-split2
     
     & ImageNet  
     & 35c-3p-34n
     & 69.85  
     & 6.11M
     & 0.98G
     
     \\
     ResNet-18-split4
     
     & ImageNet
     & 69c-5p-68n
     & 68.07    
     & 3.32M
     & 0.55G
     
     \\
     ResNet-18-split8
     
     & ImageNet
     & 137c-9p-136n
     & 66.76
     & 1.93M
     & 0.34G

     \\

  \bottomrule
\end{tabular}

\begin{tablenotes}
    \item[$\dagger$] fc: fully-connected, c: convolution, p: pooling, n: normalization, and d: dropout.
    \item[*] Detailed results are removed for brevity, refer to Figure~\ref{fig:acc-vision}. The results follows the same trend.
  \end{tablenotes}
  
\label{table:models-vision}
\vspace{-5pt}

\end{threeparttable}

\end{adjustbox}

\end{table}
\renewcommand{\arraystretch}{1}

%

\begin{figure}[t]
  \vspace{-0pt}
  \centering
  \includegraphics[width=1.0\linewidth]{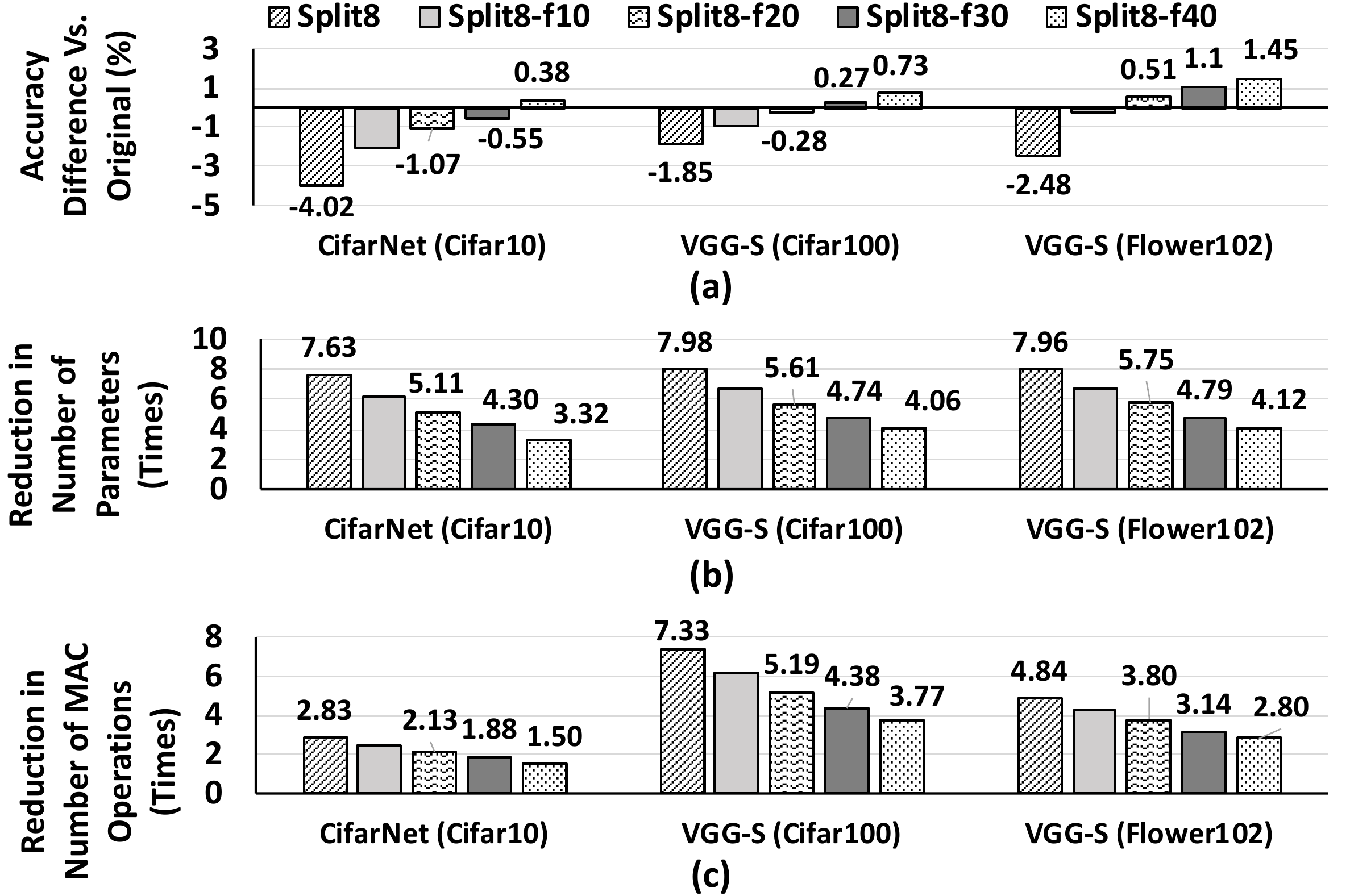}
  \vspace{-15pt}
  \caption{\textbf{Split-Fattened Models} -- Common visual models (a) Accuracy difference, (b) reduction in the number of parameters, and (c) reduction in the number of MAC operations in comparison with the original one (Table~\ref{table:models-vision}).}
  \label{fig:acc-vision-fat}
  \vspace{-0pt}
\end{figure}

%
%-----------------------------
\noindent
\textbf{Split-Fattened Models}
Accuracy is a defining factor in several applications. Thus, we provide a remedy to restore the accuracy of split-only models. By fattening (\ie, adding more parameters) each branch, we aim to create larger layers in the split-only models. To do so, for each layer (excluding classification layer) in every branch, we increase the width by a fraction. So, fattening by 20\% means the size of the output in each layer is increased 1.2$\x$. We fatten every branch in 10\% steps as Procedure~\ref{algo:split} shows. Our experiments focus on split8, which have the highest accuracy drops. Figure~\ref{fig:acc-vision-fat} shows a summary of these models. As seen, 40\% split-fattened models have higher accuracy than the original model while having fewer parameters and MAC operations. On average (for 30\% and 40\% models), with 4.61$\x$--3.81$\x$ fewer parameters and 2.95$\x$--2.5$\x$ fewer MAC operations, split-fattened models achieve accuracy within our error bound of 3\%, $\text{Task}_\text{error}$, while they jointly optimize memory, computation, and communication for edge.

%

\begin{figure}[b]
  \vspace{-0pt}
  \centering
  \includegraphics[width=1.0\linewidth]{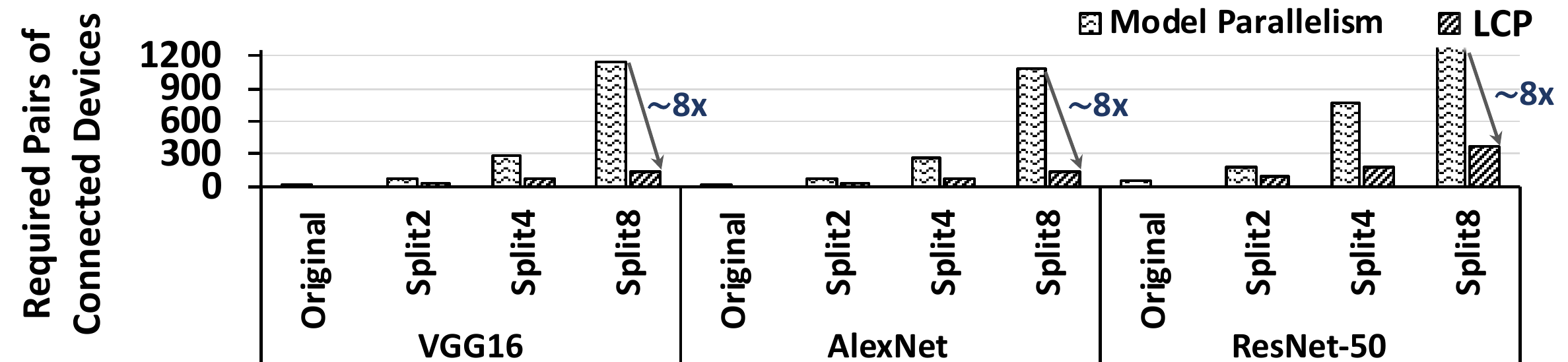}
  \vspace{-10pt}
  \caption{Communication reduction with LCP models compared to model parallelism (required pairs of connections).}
  \vspace{-5pt}
  \label{fig:communication}
\end{figure}

%
%-----------------------------
\noindent
\textbf{ImageNet Models:}
Table~\ref{table:models-imagenet-fat} illustrates the results of ImageNet models. For the sake of brevity, we only show split8 and one fattened model.  As shown, \texttt{f40} models restore the accuracy within 3\% of the original model. The tradeoff for 3\% accuracy loss is about 4$\x$ fewer parameters, 4$\x$ fewer computations, and 8$\x$ less communication load (vs. model parallelism). Figure~\ref{fig:communication} presents a comparative analysis for the communication load between distributed original models with model parallelism and distributed LCP models. Since LCP models avoid communication between their branches, the communication load is reduced significantly. In short, although split models are more complex than the original models in terms of the number of layers and connections, they achieve more parallelism with less communication load.

%

%-----------------------------
%-----------------------------
%-----------------------------
\subsection{Exploring Performance on RPis, PYNQs, and AWS}
\label{sec:res:rpi}

%----------------------
\noindent
\textbf{RPi Experiments Setup:}
To study the benefits of LCP models versus only model-parallelism methods, we deploy several models on a distributed system of Raspberry Pi 3s (RPis), the specifications in Table~\ref{tab:pi}. On each RPi, with the Ubuntu 16.04 operating system, we use TensorFlow~\cite{tensorflow2015-whitepaper} and Apache Avro~\cite{apache}, a remote procedure call (RPC) and data serialization framework, for communication between RPis. We measure power using a USB digital multimeter~\cite{UM25C}. A local WiFi network with the measured bandwidth of 62.24\,Mbps and a measured client-to-client latency of 8.83\,ms for 64\,B is used. All the real-world experiments are full-system measurements with all overheads included without any simulations/estimations.

%
% >>>>>>>>>>>>>>>>>>>>>>>>>>
% >>>>>>>>>>>>>>>>>>>>>>>>>>
% >>>>>>>>>>>>>>>>>>>>>>>>>>
%
%
%
\renewcommand{\arraystretch}{0.70}
\begin{table}[t]

\centering
\begin{adjustbox}{width=\columnwidth}

\begin{threeparttable}
\vspace{0pt}
\centering

\scriptsize
\vspace{-5pt}
\centering
\caption{Results of ImageNet LCP models.\vspace{-5pt}}
\vspace{0pt}
\begin{tabular}{c || c | c | c | c | c }
  \toprule
   \multirow{2}{*}{\textbf{Model Name}} 
      & \multirow{2}{*}{\textbf{Dataset}}
      & {\textbf{Top-1}}
      & {\textbf{Top-5}}
      & \textbf{\#} 
      & \textbf{\# MAC} \\

      & 
      & {\textbf{Acc.}}
      & {\textbf{Acc.}}
      & \textbf{Param.}
      & \textbf{MAC Opr.}\\
      \midrule 
  
     %-------------------------------------------
     \textbf{AlexNet}
     
     & \textbf{ImageNet}
     & \textbf{57.02}
     & \textbf{80.32}
     & \textbf{50.3M}
     & \textbf{678.97M}
     
     \\
     AlexNet-split8
     
     & ImageNet
     & 49.03
     & 73.10
     & 6.32M 
     & 145.37M
     
%     \\
%     AlexNet-split8-f10
%     
%     & ImageNet
%     & 51.81
%     & 74.20
%     & 7.56M 
%     & 170.1M
%
%     \\
%     AlexNet-split8-f20
%     
%     & ImageNet
%     & 52.65
%     & 74.92
%     & 8.92M
%     & 197.94M
%     
%     \\
%     AlexNet-split8-f30
%     
%     & ImageNet
%     & 53.45
%     & 75.32
%     & 10.52M  
%     & 213.04M
     
     \\
     AlexNet-split8-f40
     
     & ImageNet
     & {54.68} %{\bf \color{PineGreen} 54.68}
     & 77.06
     & 12.11M
     & 244M

     \\
     \midrule
     %-------------------------------------------
     \textbf{VGG16}
     
     & \textbf{ImageNet}
     & \textbf{70.48}
     & \textbf{90.02}
     & \textbf{138.36M}
     & \textbf{15.47G}
     
     \\
     VGG16-split8
     
     & ImageNet
     & 58.67
     & 81.54
     & 7.64M
     & 2.01G
     
%     \\
%     VGG16-split8-f10
%     
%     & ImageNet
%     & 60.02
%     & 84.12
%     & 20.77M
%     & 2.44G
%     
%    \\
%     VGG16-split8-f20
%     
%     & ImageNet
%     & 61.92
%     & 84.95
%     & 24.83M
%     & 2.89G
%     
%     \\
%     VGG16-split8-f30
%     
%     & ImageNet
%     & 64.22
%     & 87.54
%     & 29.01M
%     & 3.35G
     
     \\
     VGG16-split8-f40
     
     & ImageNet
     & 67.24 %{\bf \color{PineGreen}67.24}
     & 89.23
     & 33.78M
     & 3.87G

     \\
     \midrule
     
     %-------------------------------------------
     \textbf{ResNet-50}
     
     & \textbf{ImageNet}
     & \textbf{75.4}
     & \textbf{93.1}
     & \textbf{22.80M}
     & \textbf{4.87G}
     
     \\
     ResNet-split8
     
     & ImageNet
     & 61.79
     & 81.22
     & 5.42M
     & 0.88G
     
%     \\
%     ResNet-split8-f10
%     
%     & ImageNet
%     & 
%     & 
%     & 
%     & 
%     
%    \\
%     ResNet-split8-f20
%     
%     & ImageNet
%     & 
%     & 
%     & 
%     & 
%     
%     \\
%     ResNet-split8-f30
%     
%     & ImageNet
%     & 
%     & 
%     & 
%     & 
     
     \\
     ResNet-split8-f40
     
     & ImageNet
     & 72.12
     & 92.19
     & 8.60M
     & 1.18G

     \\
     \midrule

     %-------------------------------------------
     \textbf{MobileNet}
     
     & \textbf{ImageNet}
     & \textbf{71.7}
     & \textbf{90}
     & \textbf{4.24M}
     & \textbf{4.86G}
     
     \\
     MobileNet-split8
     
     & ImageNet
     & 59.68
     & 83.23
     & 1.12M
     & 0.93G
     \\

     MobileNet-split8-f40
     
     & {ImageNet}
     & {68.05}
     & {89.12}
     & {2.12M}
     & {1.34G}
     \\

  \bottomrule
\end{tabular}

\begin{tablenotes}
    \item[] For \texttt{[model\_name]-f[number]}, number represent the percentage of fattening.

  \end{tablenotes}

\label{table:models-imagenet-fat}

\end{threeparttable}

\end{adjustbox}

\vspace{-0pt}
\end{table}
\renewcommand{\arraystretch}{1}
%
%
%
% >>>>>>>>>>>>>>>>>>>>>>>>>>
% >>>>>>>>>>>>>>>>>>>>>>>>>>
% >>>>>>>>>>>>>>>>>>>>>>>>>>

%

%>>>>>>>>>>>>>>>>>>>
%>>>>>>>>>>>>>>>>>>>
%>>>>>>>>>>>>>>>>>>>
\begin{figure*}[t]
  \vspace{-0pt}
  \centering
  \includegraphics[width=1.0\linewidth]{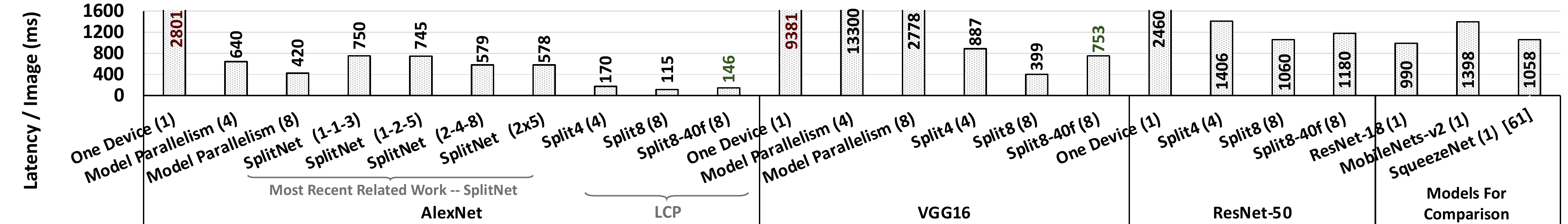}
  \vspace{-10pt}
  \caption{Latency per image: Model-parallelism, SplitNet~\cite{kim2017splitnet}, and LCP models on RPi (number in parenthesis is \#devices).}
  \vspace{-0pt}
  \label{fig:basei-rpi-top}
\end{figure*}
%>>>>>>>>>>>>>>>>>>>
%>>>>>>>>>>>>>>>>>>>
%>>>>>>>>>>>>>>>>>>>

%
%
\renewcommand{\arraystretch}{0.99}
\begin{table}[b]
    \scriptsize
	\centering

	\vspace{-5pt}
	\caption{Specification of RPi, PYNQ FPGA, and AWS.}
	\vspace{-5pt}

	%%% RPi
	\begin{tabular}{c | c | c}
		\toprule
		\multicolumn{3}{c}{\footnotesize{ \bf Raspberry Pi 3B+}} \\
		\midrule
        CPU & \multicolumn{2}{c}{1.2\,GHz Quad Core ARM Cortex-A53} \\
        Memory &  \multicolumn{2}{c}{1\,GB LPDDR2 SDRAM \textbf{@ 933Mb/s/pin}} \\
        Die Size &  \multicolumn{2}{c}{$\approx 196 {mm}^2$ @ \textbf{28\,nm}} \\
		\midrule
	\end{tabular}
	%%% Edge FPGA
	\begin{tabular}{c | c | c | c | c}
        \multicolumn{5}{c}{\footnotesize{ \bf Edge FPGA} \tiny{(Zynq Artix 7 XC7Z020)}} \\
        \midrule
        \multirowcell{3}{Utilization} &        & DSP48E & FF    & LUT  \\
                                      \cline{2-5}
                                      &\#Unit  & 96     & 5427  & 2343 \\
                                      \cline{2-5}
                                      & \%     & 44     & 5     & 4    \\
                                      \cline{1-5}
        Static Power                  & \multicolumn{4}{c}{0.121\,W}     \\
        Dynamic Power                 & \multicolumn{4}{c}{Signals: 0.009\,W~~~ Logic: 0.003\,W}     \\
        
		\midrule
	\end{tabular}
	%%% ASIC
	\begin{tabular}{c | c | c}
        \multicolumn{3}{c}{\footnotesize{ \bf AWS}} \\
        \midrule
        AWS Instance & \multicolumn{2}{c}{T2.micro} \\
        Specification &  \multicolumn{2}{c}{1 vCPU, 1\,GB Memory, 64\,GB Storage} \\
		\bottomrule
	\end{tabular}

	\label{tab:pi}
	\vspace{-5pt}
\end{table} \renewcommand{\arraystretch}{1}
%
%

%----------------------
\noindent
\textbf{RPi Performance \& Energy:}
\begin{figure}[t]
  \vspace{-0pt}
  \centering
  \includegraphics[width=1.0\linewidth]{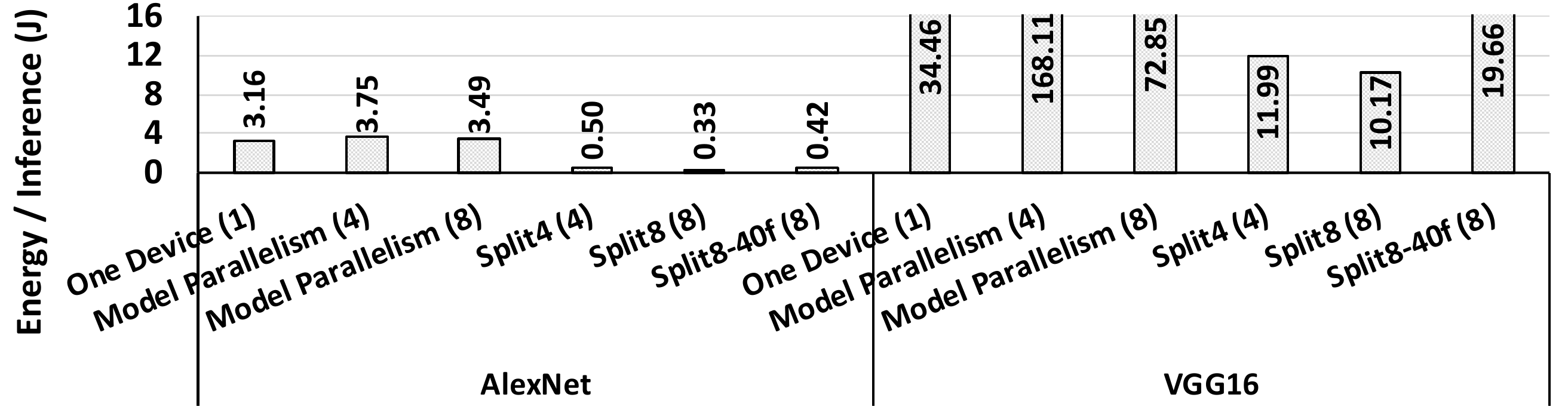}
  \vspace{-10pt}
  \caption{All devices energy per inference: Model-parallelism, and LCP on RPi (number in parenthesis is \#devices).}
  \vspace{-00pt}
  \label{fig:basei-rpi-energy}
\end{figure}
Figure~\ref{fig:basei-rpi-top} presents latency of inference per image on RPis. On a single device, AlexNet has 2.8 seconds latency, while VGG16 achieves 9.4 seconds latency. By deploying model-parallelism variants of the models on four and eight RPis, we achieve a maximum of 0.42s latency, a 6.6$\x$ increase, for AlexNet. But, for VGG16, on four RPis, we observe a slowdown, which is caused by high communication latency.  LCP variants of split4 and split8 can reach up to 115\,ms and 400\,ms latency per image for AlexNet and VGG16, respectively. This is because LCP models are lightweight and parallelizable and have low communication. Figure~\ref{fig:basei-rpi-energy} shows measured energy per inference for RPi implementations.
To compare with previous related work, SplitNet~\cite{kim2017splitnet}, Figure~\ref{fig:basei-rpi-top} presents the performance of SplitNet models for AlexNet with different configurations. As seen, the performance is worse than LCP models. This is because SplitNet creates more merging/synchronization points with its tree-structured model design. The resulting model exponentially introduces more merging/synchronization with increased depth, which also does not equally split all the layers (causing load balancing issues). Finally, SplitNet performs parallelization based on dataset semantics, which means every dataset and model needs to be manually split. \cref{sec:motive} provided more reasons on this performance difference.

%----------------------
\noindent
\textbf{TVM Experiments on PYNQ Boards:} As a real-world example for edge FPGA implementation, we use TVM~\cite{chen2018tvm} on the PYNQ~\cite{pynq} board. PYNQ is designed for embedded applications. We use the TVM VTA stack on the PYNQ as the architecture (RISC-style instructions) and only change the models (ResNet-18 vs. LCP ResNet-18 Split2 with $<\!\!1$ accuracy drop). In this way, we can measure the benefits of LCP models without relying on any special tailored hardware.
\begin{wrapfigure}{r}{0.23\textwidth}
  \begin{center}
  \vspace{-20pt}
    \hspace{-20pt}\includegraphics[width=0.23\textwidth]{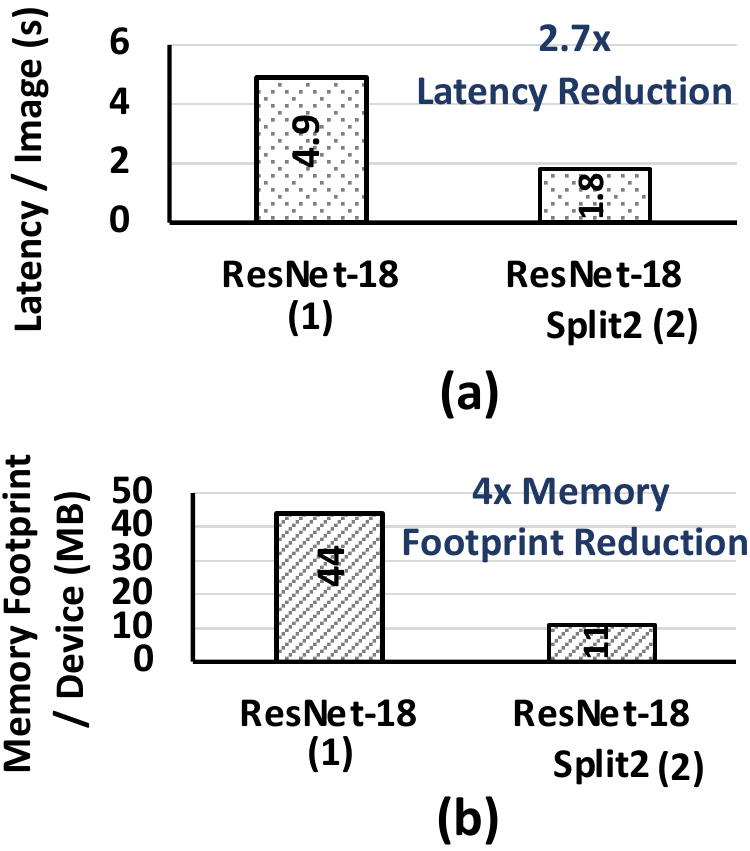}
  \end{center}
  \vspace{-12pt}
   \caption{TVM Experiments: (a) Latency per image, (b) memory footprint per device (number in parenthesis is \#devices).}
  \vspace{-12pt}
  \label{fig:table-pynq}
\end{wrapfigure}
Our performance result shares the entire system pipeline performance, from a live camera feed to prediction output on two boards versus one board. Figure~\ref{fig:table-pynq}a shows a 2.7$\x$ speedup, including all communication and system overheads, network latency, and jitter because LCP models are parallelized on two devices and, in total, they have lower computation and memory footprints. The measured reduction in memory footprint is shown Figure~\ref{fig:table-pynq}b.
\begin{figure}[b]
  \vspace{-0pt}
  \centering
  \includegraphics[width=1.0\linewidth]{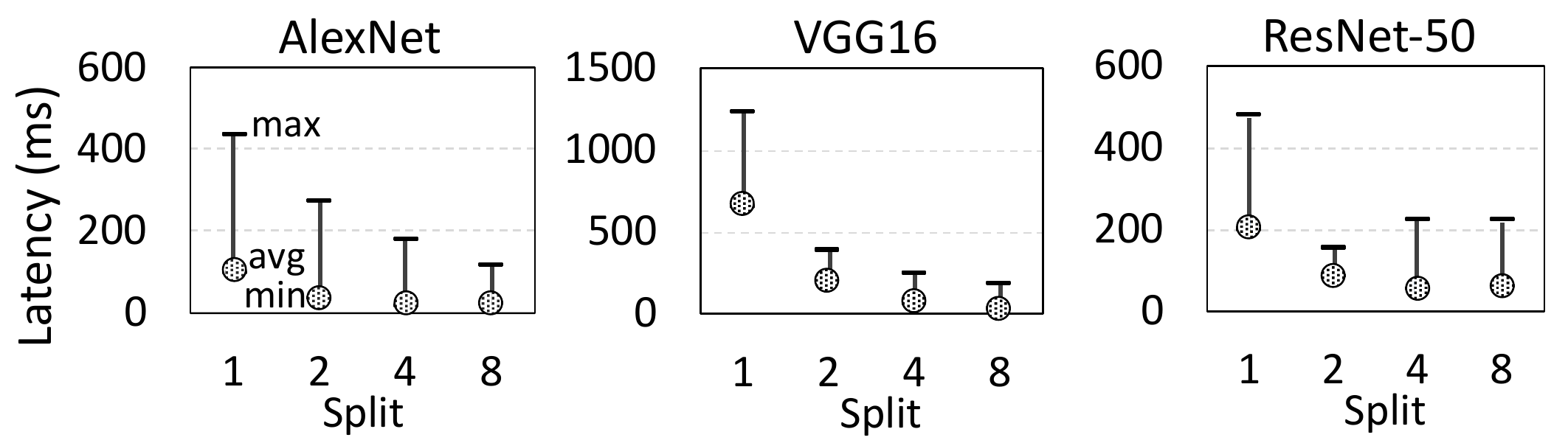}
  \vspace{-15pt}
  \caption{Average, minimum, and maximum latencies of distributed LCP execution on AWS T2.micro instances with 1 vCPU and 1\,GB memory per instance.}
  \vspace{-0pt}
  \label{fig:aws}
\end{figure}
%----------------------
\noindent
\textbf{AWS Experiments:} To see the reduced communication and distributed execution benefits of LCP models further, we deploy AlexNet, VGG16, and ResNet-50 models on AWS T2.micro instances with only one vCPU and 1\,GB memory per instance. Figure~\ref{fig:aws} presents the derived statistics. In all cases, LCP models not only reduces the average latency but also significantly reduce maximum latency. Splits four and eight have lower speedup compared with our RPi experiments because all the 4/8 instances are not hosted on the same machine; thus, the communication cost is higher than the usual edge-specific cases that this paper targets.

%-----------------------------
%-----------------------------
%-----------------------------
\subsection{Edge FPGA Experiments}
\label{sec:res:fpga}

\noindent
\textbf{FPGA Experiments Setup:}
We implement our tailored microarchitecture on a ZYNQ XC7Z020 FPGA targeting PYNQ-z1 boards~\cite{bae2019capella}. We use Xilinx Vivado HLS for implementation and verify the functionality of our implementation using regression tests. We use relevant \textit{\#pragrma} as hints to describe our desired microarchitectures in C++. We synthesize and implement our design using Vivado and report post-implementation (\ie, place \& route) performance numbers and resource utilizations. Inputs and output of our design are transferred through the AXI stream interface. The clock frequency is set to 100\,MHz. Communication for multiple devices is estimated with the network provided in~\cref{sec:res:rpi}.

\begin{figure}[b]
  \vspace{-0pt}
  \centering
  \includegraphics[width=1.0\linewidth]{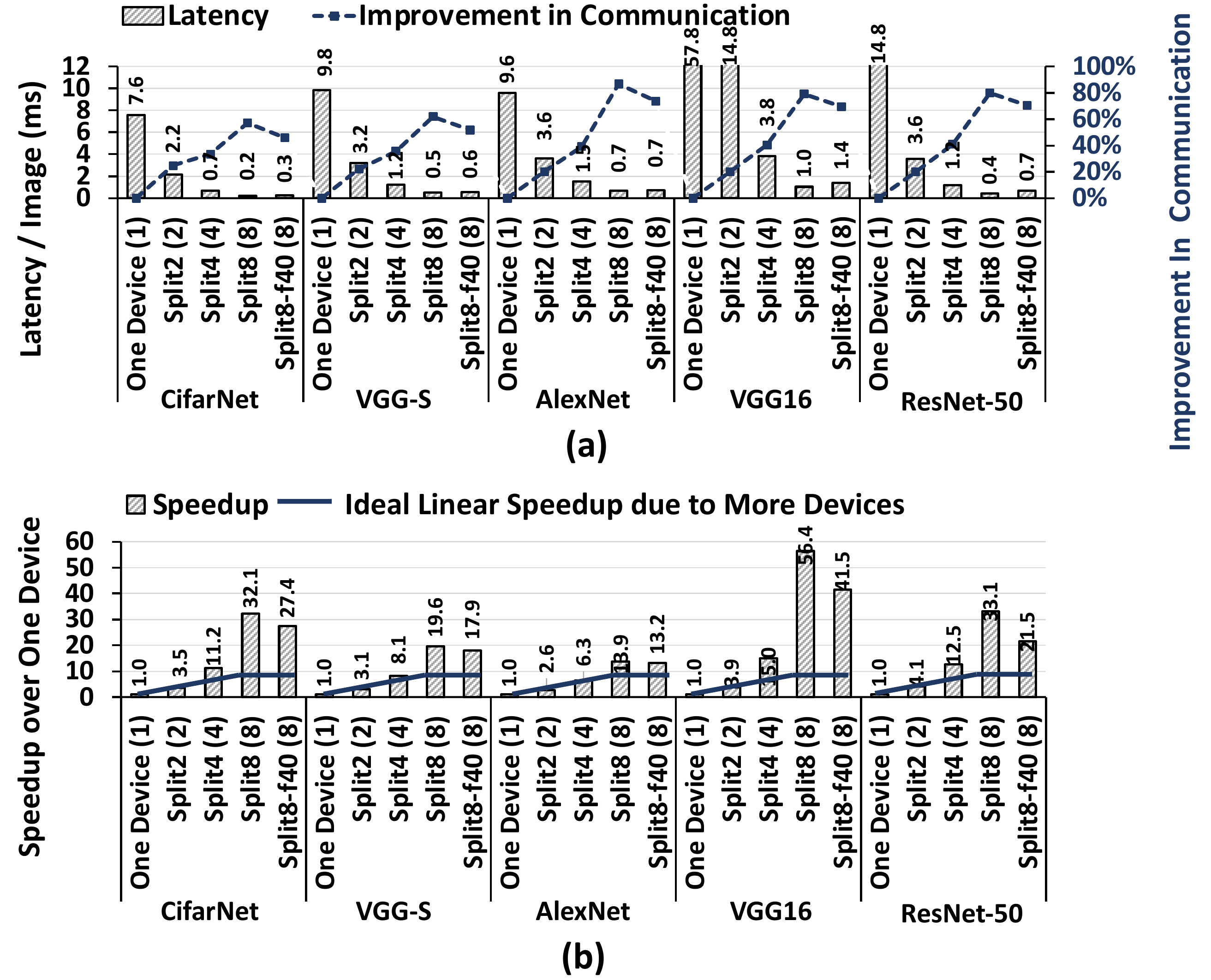}
  \vspace{-15pt}
  \caption{Edge FPGA with tailored hardware latency and speedup: (a) Latency per image, (b) speedup over one device (number in parenthesis is \#devices).}
  \vspace{-0pt}
  \label{fig:base-uetp}
\end{figure}
%

%>>>>>>>>>>>>>>>>>>>
%>>>>>>>>>>>>>>>>>>>
%>>>>>>>>>>>>>>>>>>>
\begin{figure*}[t]
  \vspace{-0pt}
  \centering
  \includegraphics[width=1.0\linewidth]{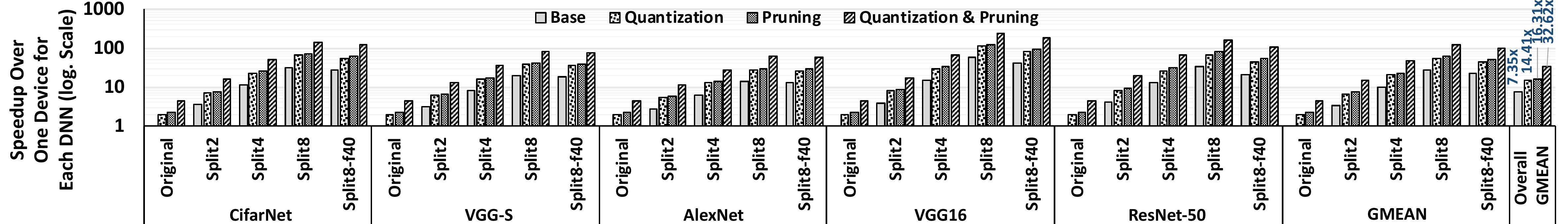}
  \vspace{-10pt}
  \caption{Edge FPGA with tailored hardware speedup with quantization \& pruning. Additional speedup is gained by applying lossless ($\leq\!0.1\%$) quantization and structured pruning.}
  \vspace{0pt}
  \label{fig:prunings}
\end{figure*}
%>>>>>>>>>>>>>>>>>>>
%>>>>>>>>>>>>>>>>>>>
%>>>>>>>>>>>>>>>>>>>

\begin{figure}[t]
  \vspace{5pt}
  \centering
  \includegraphics[width=1.0\linewidth]{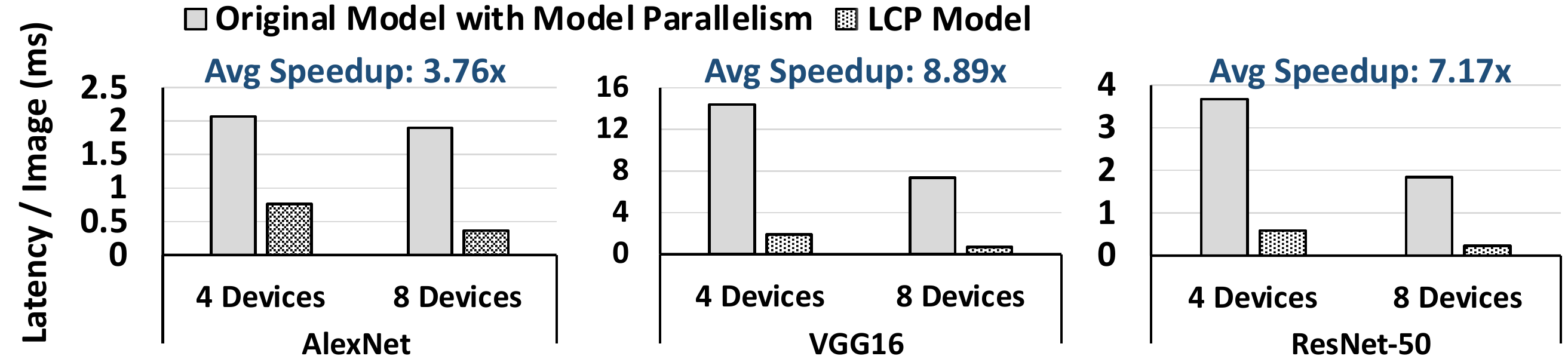}
  \vspace{-10pt}
  \caption{Latency per image for edge FPGA with tailored hardware comparing LCP vs. model parallelism.}
  \vspace{-5pt}
  \label{fig:base-uetp2}
\end{figure}

%----------------------
\noindent
\textbf{FPGA Performance:}
Figure~\ref{fig:base-uetp} shows the experiment results for our edge-tailored hardware. The latency per image is shown in Figure~\ref{fig:base-uetp}a, with improvement in communication overhead versus model-parallelism methods (86\% and 60\% for 8split and 4split). Depending on the model, the inference per latency on a single device is between 4--29ms; a 221--325$\x$ speedup compared to RPi results for AlexNet and VGG16. Our designed LCP models achieve acceptable performance for edge computing, which is 10s of inferences per second, around 1--10ms. As observed, the accuracy loss of our split-only models can be easily restored by fast split-fattened models of \texttt{f40} with a negligible performance overhead (maximum of 20\,ms). Figure~\ref{fig:base-uetp}b illustrates the speedup numbers over one device. The ideal linear speedup shows the ideal scaling speedup with more available devices. As shown, we achieve superlinear speedups. An important parameter in scaling concerns how the \textit{overheads} scale. The superlinear speedup stems from the dramatic reduction of communication overhead as parallelism increases. In traditional data and model parallelism, such overhead increases, which causes sublinear speedup.
Figure~\ref{fig:base-uetp2} compares latency per image for LCP and model parallelism. On average, LCP models are 3.76$\x$, 8.89$\x$, and 7.17$\x$ faster than their model-parallelism counterparts for AlexNet, VGG16, and ResNet-50 (4 and 8 devices), respectively. LCP achieves a maximum and average speedups of $56\x$ and $7\x$, compared to the originals (Figure~\ref{fig:prunings}, base bars).

%

%----------------------
\noindent
\textbf{Quantization \& Pruning:}
As mentioned in \cref{sec:related} and \cref{sec:intro}, techniques that reduce the footprint of DNNs can be applied to each individual LCP branch. Basically, the target output for each LCP branch is now its pre-final activations during optimizations. We study the benefits of lossless quantization and structured pruning on top of our LCP models. Based on our experiment, with 3.13 ($<$integer.fraction$>$) quantization, our models do not lose accuracy. Similarly, applying structured pruning~\cite{anwar2017structured}, for which systolic arrays gain benefits, reduces the size of parameters between 40\%--50\% per convolution layer without an accuracy drop. Other pruning algorithms increase the sparsity of the data, which is not necessarily beneficial for systolic arrays. Figure~\ref{fig:prunings} presents the speedup gained from these techniques normalized to the baseline implementation for each model, the execution performance of which shown in Figure~\ref{fig:base-uetp}a. Quantization and pruning themselves, improve the performance of the original models by 1.96$\x$ and 2.2$\x$, respectively, and 4.31$\x$ when applied together. When quantization and pruning are combined with LCP, the overall performance speedup becomes 14.41$\x$ and 16.31$\x$, respectively. Compared to the original models, LCP + quantization and pruning achieves up to 244$\x$ speedup (VGG16-split8), and an average of 33$\x$ (across all models and variants).

%
%-----------------------------
%-----------------------------
%-----------------------------
\subsection{ASIC Implementation}
\label{sec:res:asic}

\begin{figure}[b]
  \vspace{-0pt}
  \centering
  \includegraphics[width=0.85\linewidth]{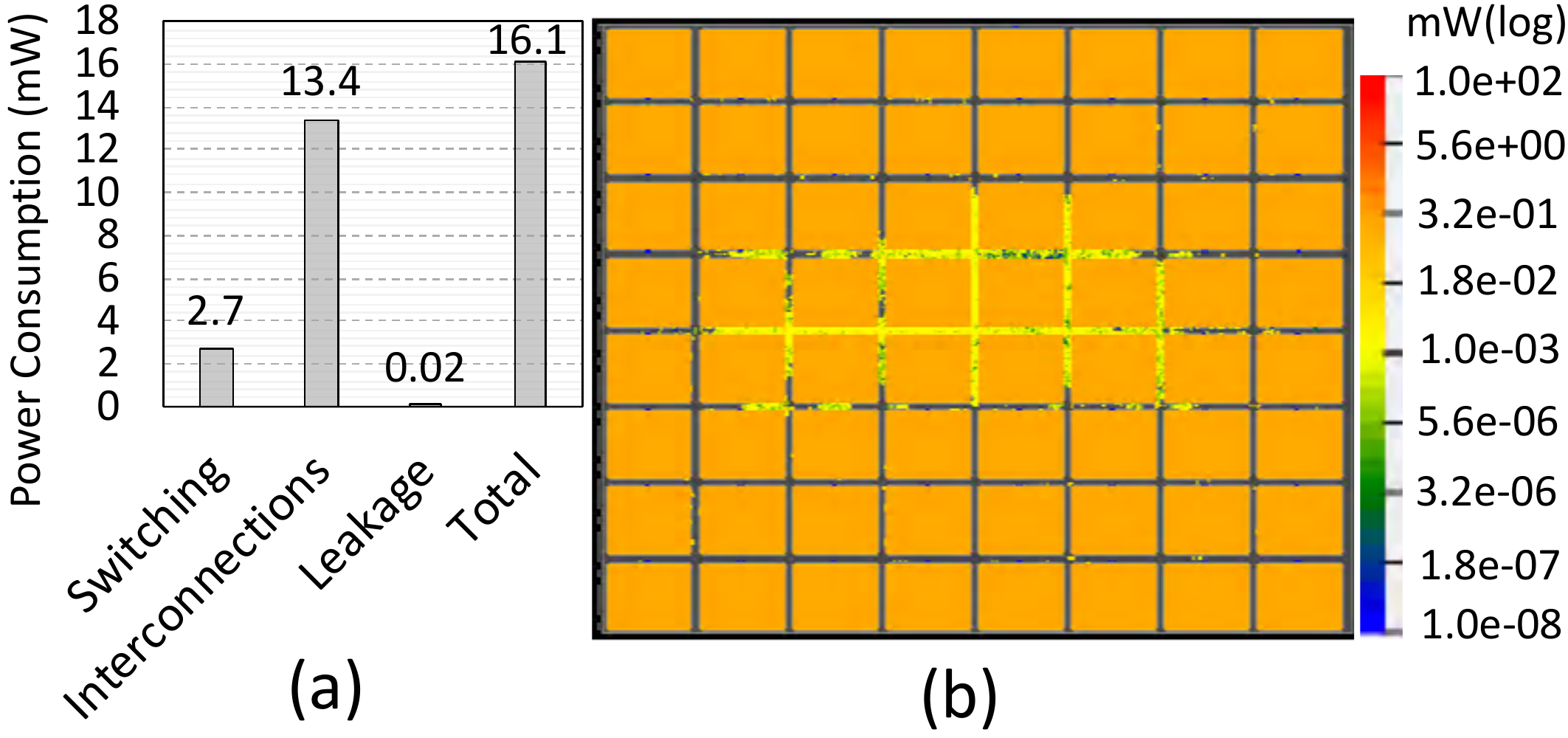}
  \vspace{-0pt}
  \caption{Power Consumption for 7-nm ASIC Design @800MHz: (a) breakdown (b) distribution.}
  \vspace{-0pt}
  \label{fig:base-uetp-layout}
\end{figure}

\noindent
We implement the ASIC design of LCP using an Arizona State Predictive PDK (ASAP) 7nm technology node~\cite{clark2016asap7}. Our tool chain includes the Synopsys design compiler (DC) for synthesis, Cadence Innovus for place and route, and Cadence Tempus for timing and power analysis. As an input to our ASIC design, we use our same Verilog code generated by Vivado HLS. Figure~\ref{fig:micro}b show the layout of our chip of size 0.107~mm$^2$ (i.e., 295$\mu m \times$ 365$\mu m$). The memory cells shown in the figure represent the FIFO buffers, used for pipelining. Figure~\ref{fig:base-uetp-layout} shows the power consumption of our ASIC design. The breakdown of power consummation leading to a total 16.1~mW is listed in Figure~\ref{fig:base-uetp-layout}a. As a comparison point, Eyeriss~\cite{eyeriss} and EIE~\cite{han2016eie} consume $\approx$250\,mW and $\approx$590\,mW, respectively. Besides, as Figure~\ref{fig:base-uetp-layout}b shows, power distributes uniformly, which prevents hot spot creation.

%%---------------------------------
\section{Related Work}
\label{sec:related}
\noindent
We review related techniques used to reduce the high demands of DNNs, distributing their computation, and current efforts on DNN hardware accelerators. 

%-----------------------------------------
\noindent
\textbf{Techniques Without Changing Model Architecture:}
Several techniques have been developed to reduce the computation and memory footprint of DNNs without changing the network architecture. For instance, pruning~\cite{yu:luk17,han:mao15,lin2017runtime, wen:wu16, asgari2019lodestar} removes the close-to-zero weights and quantization or low-precision inference~\cite{cou:ben14, gon:li14, van:sen11,koster2017flexpoint,lin2016fixed} change the representations of numbers, which results in simpler calculations. Other methods partition resources~\cite{she:fer16, guo:yin17} or binarize the weights~\cite{li:zha16, cou:hub:16, rast:ord16}. Binarizing weights hurts accuracy. \emph{The aforementioned techniques are orthogonal to our work and can be applied to each branch to further reduce the computational and memory costs (\cref{sec:res:fpga}).}

%-----------------------------------------
\noindent
\textbf{Techniques That Change Model Architecture:}
With the prevalence of IoT and edge devices, specific frameworks such as ELL library~\cite{ell} (see Figure~\ref{fig:rpi-times}) by Microsoft and Tensorflow Lite~\cite{tensorflowLite} have been developed by industry. Other proposals developed mobile-specific models~\cite{how:zhu17, ian:mos16, shufflenet, condensenet, mobilenetv2} by handcrafting more efficient operations or models to reduce the number of parameters~\cite{ian:mos16}, create efficient operation to minimize computation density~\cite{how:zhu17}, or use resource-efficient connections~\cite{mobilenetv2}. Unlike LCP models, all these models have a single chain of dependency~\cite{xie2019exploring} that prevents efficient parallelism. Moreover, several of the models trade off the state-of-the-art accuracy with efficiency~\cite{mobilenetv2}. SplitNet~\cite{kim2017splitnet} is one the few papers that focuses on higher parallelizability of models (evaluation on ResNet and AlexNet), but relying on dataset semantics creates imbalanced branches and the method is invariant to number of devices, as discussed in \cref{sec:motive}. 
Recently, with the growing interest in automating the design process~\cite{nasnet1, nasnet2, metaQNN, xie2019exploring}, learning new networks for mobiles has also gained attention by integrating the constraints of mobile platforms (\ie, latency). These attempts are still limited to single-device execution.
\emph{In summary, these studies (1) have a high design cost, (\ie, they target only one specific model and dataset without extendibility); (2) target single mobile platforms; and (3) do not consider inter-layer layer parallelism and communication challenges.}

%-----------------------------------------
\noindent
\textbf{Distributing DNN Inference Computations:}
With large DNN models, distributing a single model has gained the attention of researchers~\cite{dean:cor12, tee:mcd17, kan:hau17, hadidi2018distributed, hadidi2020towards}. Usually, the distribution is done in a high-performance computing domain with different goals in mind. In the resource-constrained edge devices, Neurosurgeon~\cite{kan:hau17} dynamically partitions a DNN model between a \emph{single} edge device and the cloud. DDNN~\cite{tee:mcd17} partitions the model between edge devices and the cloud but uses data parallelism. Hadidi et al.~\cite{hadidi2018distributed, hadidi2018musical, hadidi2020towards, cao2019edge, hadidi2018real, hadidi2019collaborative} investigate the distribution in edge with model-parallelism methods, showing the effect of the communication barrier in distributing by the diminishing return in performance with a large number of devices. \emph{LCP models go beyond model parallelism methods, which was not the focus of the above studied, and enable efficient distribution that is not examined in the above studies.}

%-----------------------------------------
% \noindent
% \textbf{Edge-Targeted and Systolic-Based Hardware:}
% Several studies explored in/near-the-edge DNN computation without proposing new hardware~\cite{kan:hau17, lei:sen13, mat:des17, lane:bha17, lane:geo15, han:mao15} by training new models, proposing collaboration techniques with cloud, or applying several device-specific and model-specific techniques. A state-of-the-art systolic-based DNN accelerator is TPUv2~\cite{dean2017machine}, which provides a peak of 180\,TFLOPS by employing four dual-core chips, each connected to an 8GB HBM package at 300\,GB/s. Many other recent deep-learning accelerators utilize systolic arrays concepts~\cite{Ese, eyeriss, shidiannao, tpu, dean2017machine}, which increase the performance of inference by utilizing sparsity, reducing memory accesses by exploring access patterns, or employing weight-stationary architectures. Several studies target FPGA/ASIC implementations for DNNs~\cite{sharma2016high, zhang2015optimizing, suda2016throughput, qiu2016going}. \emph{These papers study the execution of the entire model on a device with no resource constraints, whereas our focus is enabling the inference distribution on several devices.}

% CITE systolic asgari2019eridanus
% CITE FPGA bae2019capella reagen2016minerva

%%---------------------------------
\section{Discussions}
\label{sec:diss}
\noindent
%
%-----------------------------
\noindent
\textbf{Intuition Behind LCP:}
We conjecture that LCP models provide good performance because (1) independent branches learn complex non-overlapping features independently within a small search space, whereas original models need to create the same complex features from a higher dimension feature search. We observe that each branch eventually learns an almost disjoint feature representation; (2) In split models,  gradient descent updates are more efficient in reaching early layers compared to the original models due to fewer number of parameters in their route.

%-----------------------------
\noindent
\textbf{Extension to New Models:} We studied ResNets and MobileNet, which are still widely used models. Other models represent sequential DNNs that serve as the basis to confirm our method and are still used in robotics. Newer models such as EfficientNet~\cite{tan2019efficientnet} and MobileNetv3~\cite{howard2019searching} that use novel blocks such as Bottleneck or Squeeze \& Excitation can be represented with convolution, fully connected, and basic matrix multiplications. All of which can be parallelized by LCP.

%-----------------------------------------
\noindent
\textbf{System-Level Choices:}  
LCP is in conjunction with other technologies available today. LCP does not replace these technologies, but rather enables exploitation of local edge devices to enable intelligence in the edge~\cite{zhou2019edge, hadidi2019robustly}. In a few cases, relying on cloud-based offloading for accuracy-critical tasks is necessary (\eg, finding a specific license plate), whereas, in several others (\eg, counting the cars passing an intersection) the system must rely on cloud or high-performance systems.

%-----------------------------------------
\noindent
\textbf{SqueezeNet:}
SqueezeNet~\cite{ian:mos16} achieves an accuracy similar to that of AlexNet with fewer parameters by using compute-heavy Fire modules. SqueezeNet trades off parameters with computations, and requires 860M MAC operations, whereas our distributed AlexNet requires only 240M MAC operations. We also observe a 12$\x$ increase in the number of activations from 12.58\,M in SqueezeNet vs. 1.39\,M in AlexNet.
 
%-----------------------------------------
\noindent
\textbf{Skip/Residual Connections:}
LCP procedure similarly applies to more complex models with residual and skip connections as shown for ResNets in \cref{sec:res}. Simply put, each branch has similar connections but with smaller depth.

%-----------------------------------------
\noindent
\textbf{Alleviating Large Memory Footprints:}
Sometimes large memory footprints are necessary and access to the next levels of the storage system is enforced. In our design (\cref{sec:ETP:micro}), such accesses do not cause slowdown because data is stored in sequential addresses (\ie, streaming), and we overlap data transfer and computations for independent elements. 

%-----------------------------------------
\noindent
\textbf{Memory Layout Preprocessing:}
Our simple algorithm to change the storage format is in $O(N)$ (\cref{sec:ETP:micro}(4)). Therefore, the host preprocessing for reordering the data can be done during writing the data to the memory with a single pass.
%

%%---------------------------------
\section{Conclusions}
\label{sec:com}
\noindent
We proposed low-communication parallelization (LCP) models, designed for efficient in-the-edge distribution. LCP models optimize communication while reducing memory and computation by utilizing several narrow independent branches. We presented our results on the accuracy of LCP models. We build a systolic architecture for edge computing both on FPGA and ASIC. Finally, our results on RPis, edge-based FPGAs, AWS instances confirms the benefits.

%%%%%%% -- PAPER CONTENT ENDS -- %%%%%%%%

%%%%%%%%% -- BIB STYLE AND FILE -- %%%%%%%%
% \bibliographystyle{IEEEtranS}
\bibliographystyle{unsrt}
\bibliography{ms}
%%%%%%%%%%%%%%%%%%%%%%%%%%%%%%%%%%%%

\end{document}